\documentclass[aps,pra,superscriptaddress, notitlepage, twocolumn, 10pt, table, floatfix]{revtex4-1}

\usepackage[T1]{fontenc}
\usepackage{lmodern}
\usepackage{mathrsfs} 
\usepackage{graphicx}
\usepackage{bm}
\usepackage{amsmath}
\usepackage{amssymb}
\usepackage{todonotes}
\usepackage{nicefrac}
\usepackage{graphicx,booktabs,array}
\usepackage{hyperref, url}
\usepackage[caption=false]{subfig} %

\hypersetup{
    colorlinks,
    linkcolor={red!50!black},
    citecolor={blue!50!black},
    urlcolor={blue!80!black}
}

\begin{document}

\title{Probing the limits of the rigid-intensity-shift model in differential phase contrast scanning transmission electron microscopy}

\author{L. Clark}
\email[Corresponding author: ]{laura.clark@monash.edu}
\affiliation{School of Physics and Astronomy, Monash University, Victoria 3800, Australia}
\author{H.G. Brown}
\affiliation{School of Physics and Astronomy, Monash University, Victoria 3800, Australia}
\author{D.M. Paganin}
\affiliation{School of Physics and Astronomy, Monash University, Victoria 3800, Australia}\author{M.J. Morgan}
\affiliation{School of Physics and Astronomy, Monash University, Victoria 3800, Australia}
\author{T.~Matsumoto}
\affiliation{Institute of Engineering Innovation, University of Toyko, Toyko 113-8656, Japan}
\author{N. Shibata}
\affiliation{Institute of Engineering Innovation, University of Toyko, Toyko 113-8656, Japan}
\author{T.C. Petersen}
\affiliation{School of Physics and Astronomy, Monash University, Victoria 3800, Australia}\author{S.D. Findlay}
\affiliation{School of Physics and Astronomy, Monash University, Victoria 3800, Australia}

\date{\today}

\begin{abstract}
The rigid-intensity-shift model of differential phase contrast scanning transmission electron microscopy (DPC-STEM) imaging assumes that
the phase gradient imposed on the probe by the sample causes the diffraction pattern intensity to shift rigidly by an amount proportional to that phase gradient.  This
behaviour is seldom realised exactly in practice.  Through a combination of experimental results, analytical modelling and numerical calculations, we explore the breakdown of the rigid-intensity-shift behaviour and how this depends on the magnitude of the phase gradient and the relative scale of features in the phase profile and the probe size. We present guidelines as to when the rigid-intensity-shift model can be applied for quantitative phase reconstruction using segmented detectors, and propose probe-shaping strategies to further improve the accuracy.
\end{abstract}

\maketitle

\section{Introduction}

Samples in transmission electron microscopy (TEM) impart phase shifts on the electron  beam that passes through them. These phase shifts follow from the Aharonov-Bohm effect \cite{aharonov1959significance} and thus encode information on the electric and magnetic  fields within the sample.  Longer range (nm--$\mu$m) fields include those caused by magnetic domains and in-built electric fields within semiconductor devices.  Within the projection and phase object approximations \cite{vulovic, brownmeasuring} -- often valid for typical electron acceleration voltages \mbox{(80--300~keV)} and sample thicknesses (which at lower resolution might extend to $\approx$~100~nm) \cite{nellistscanning} -- the exit surface wavefunction, $\Psi_{\rm exit}$, is related to the entrance surface wavefunction, $\Psi_{\rm entrance}$, via multiplication with a transmission function, $T(x,y)$:
\begin{eqnarray}
\Psi_{\rm exit}(x, y)&=&\Psi_{\rm entrance}(x, y)\cdot T(x,y) \nonumber \\ &=&\Psi_{\rm entrance}(x, y)\cdot \exp[i \phi (x,y)],
 \label{eq:a1}
\end{eqnarray}
where $\phi(x,y)$ describes the sample-induced phase shift.  This imparted phase is lost if the exit wavefunction is imaged in focus, but a range of techniques exist that convert this imparted phase to measurable intensity changes. 

In conventional TEM, where the entrance wavefunction is a plane wave, weak phase shifts (much less than $\pi$) can be visualised using Zernike phase-contrast \cite{bornwolf, malachole, danevvolta} or out-of focus (Lorentz) imaging \cite{DeGraefZhuBook, hopstermagnetic, phatakrecent}.  For the stronger phase shifts usually pertaining to materials specimens \cite{beleggia, vulovic}, techniques include through-focal series phase retrieval \cite{C2,AMOO1} and off-axis holography \cite{RDBchapter}.  A strength of conventional TEM is synchronous acquisition of the full field of view: images are recorded with perfect registration between pixels.

\begin{figure}
    \centering
\includegraphics[width=0.95\columnwidth]{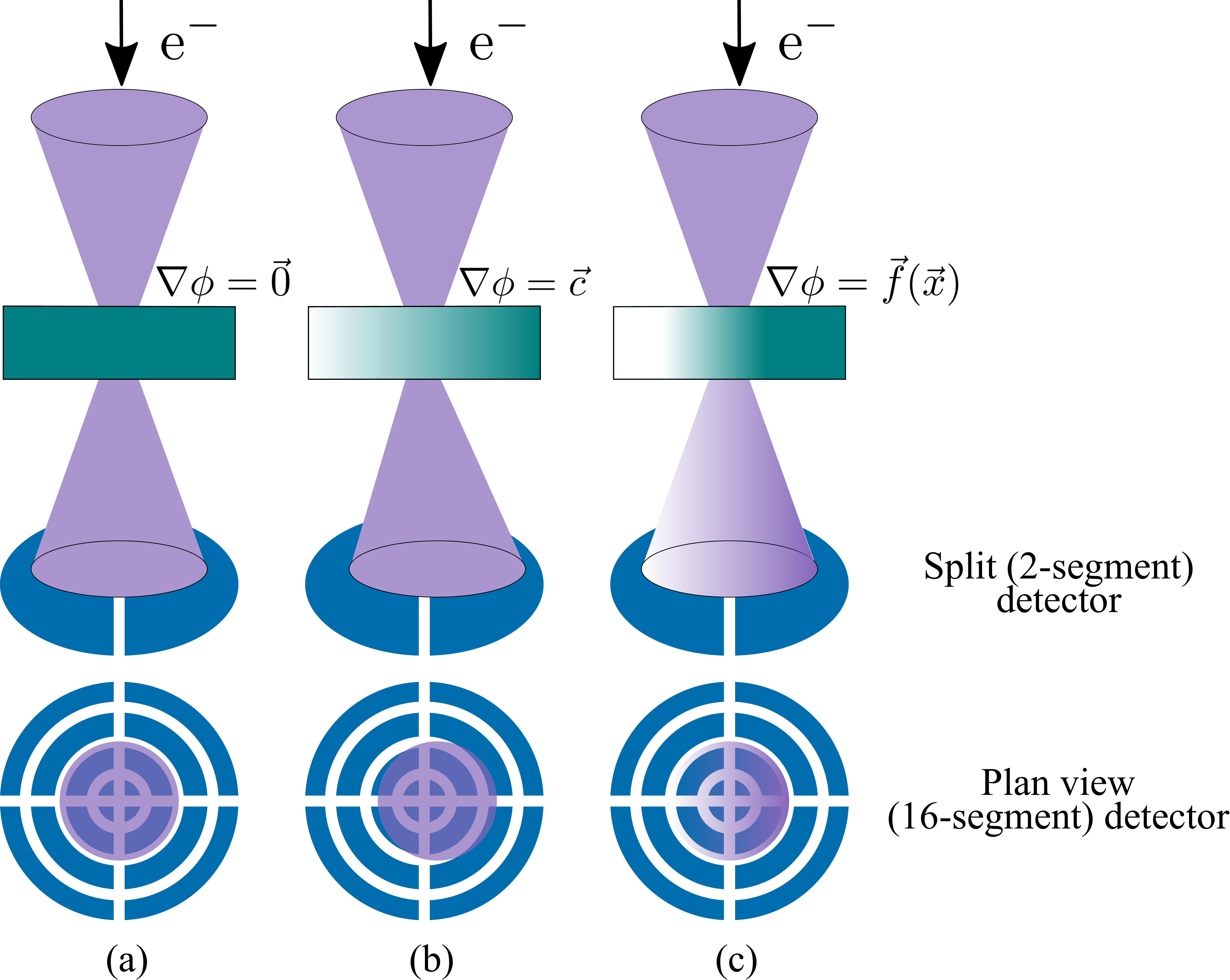}
\caption{DPC-STEM imaging. (a) When $\nabla \phi= \vec{0}$, the diffraction pattern of the probe is unshifted. (b) When $\nabla \phi = \vec{c}$, the diffraction pattern is rigidly shifted across the detector, with deflection $\beta \propto |\nabla \phi$|. (c) When $\nabla \phi$ varies across the width of the probe distribution, the diffraction pattern intensity is not simply rigidly shifted, but rather redistributed across the detector in a more complex manner.}
\label{fig:DPCschematic}
\end{figure}

In scanning-TEM (STEM), the electron beam is focussed into a small probe at the sample entrance surface by a set of condenser lenses and apertures, of which the final aperture, characterised by its convergence semiangle, $\alpha$, determines the probe size at the sample \cite{williamscarter}.  The probe is then scanned across the specimen, with the STEM image(s) built up by plotting the recorded signal(s) as a function of probe position.  This serial acquisition over probe positions can introduce image distortions, but allows for multiple images to be recorded in perfect registration to one another.  This makes STEM a powerful technique for correlative imaging.  

Fig.~\ref{fig:DPCschematic}a depicts the convergent probe incident upon a sample that imparts a constant phase shift, with the effect that the intensity distribution in the far-field (also called the diffraction plane) is an image of the probe-forming aperture, called the bright-field disk.  Fig.~\ref{fig:DPCschematic}a further depicts this bright-field disk falling on example configurations of multiple detectors (split detector in profile view; segmented detector in plan view).  
This subdivision of the detector plane
allows direct imaging of the phase gradient via differential phase contrast (DPC), as first suggested and implemented in the 1970s \cite{roseelectron, dekkersdifferential}.  The essence of this approach is shown in Figs.~\ref{fig:DPCschematic}b and c.  In classical terms the Lorentz force from an electric field that is perpendicular to the optical axis, or in wave-optical terms a transverse gradient in the imparted phase, deflects the beam laterally. If the detector comprises multiple, non-rotationally-symmetric elements, this redistribution of the intensity in the diffraction plane (the diffraction pattern) can be measured using the difference in recorded intensity on the different detector segments.

In the wave-optical formulation, the diffraction pattern intensity can be described using Eq. (\ref{eq:a1}) as:
\begin{align}
I(k_x, k_y)=&|\mathcal{F}\left[ \Psi_{\rm exit}(x,y)  \right] |^2\\
=&| \Psi_{\rm entrance}(k_x, k_y) \otimes \mathcal{F}\{\exp[i \phi (x,y)]\} |^2\,,
\label{eq:ronchiInt}
\end{align}
where $\otimes$ denotes convolution and $\mathcal{F}$ denotes Fourier transform with respect to $x$ and $y$, and $k_x$ and $k_y$ are the corresponding Fourier-space coordinates.  If the gradient of the imparted phase is strictly linear over the full width of the incident wavefunction, the Fourier shift theorem implies a rigid deflection of the diffraction pattern \cite{goodmanintroduction,lubkzweck,lazic2016phase}, as depicted in Fig.~\ref{fig:DPCschematic}b.  Conversely, when the gradient of the imparted phase varies across the incident wavefunction, the exact rigid-intensity-shift model is not expected to hold \cite{muller2014atomic,lazic2016phase,cao2017theory} and the intensity in the diffraction pattern will be redistributed in a more complex fashion, as depicted in Fig.~\ref{fig:DPCschematic}c.   
DPC imaging is often conceptualised in terms of the diffraction pattern undergoing a rigid intensity shift (also called rigid disk shift since in the absence of more complex intensity redistribution the diffraction pattern is disk shaped) -- especially so for the case of longer-range field imaging \cite{chapman1992differential,lohr2012differential,zweckdetector,krajnak2016pixelated,lohr2016quantitative,schwarzhuber2017achievable,wu2017correlative}. 
In many microstructured materials, however, the phase gradient varies across the illuminated specimen region and so the rigid-intensity-shift model may not hold exactly.  A particular example of interest, which we revisit and extend presently, was given in the p-n junction DPC-STEM imaging work of Shibata \emph{et al.} \cite{SFSMSKOMI1}.

\begin{figure}[ht]
\centering
\subfloat[][]{
\includegraphics[width=0.9\columnwidth]{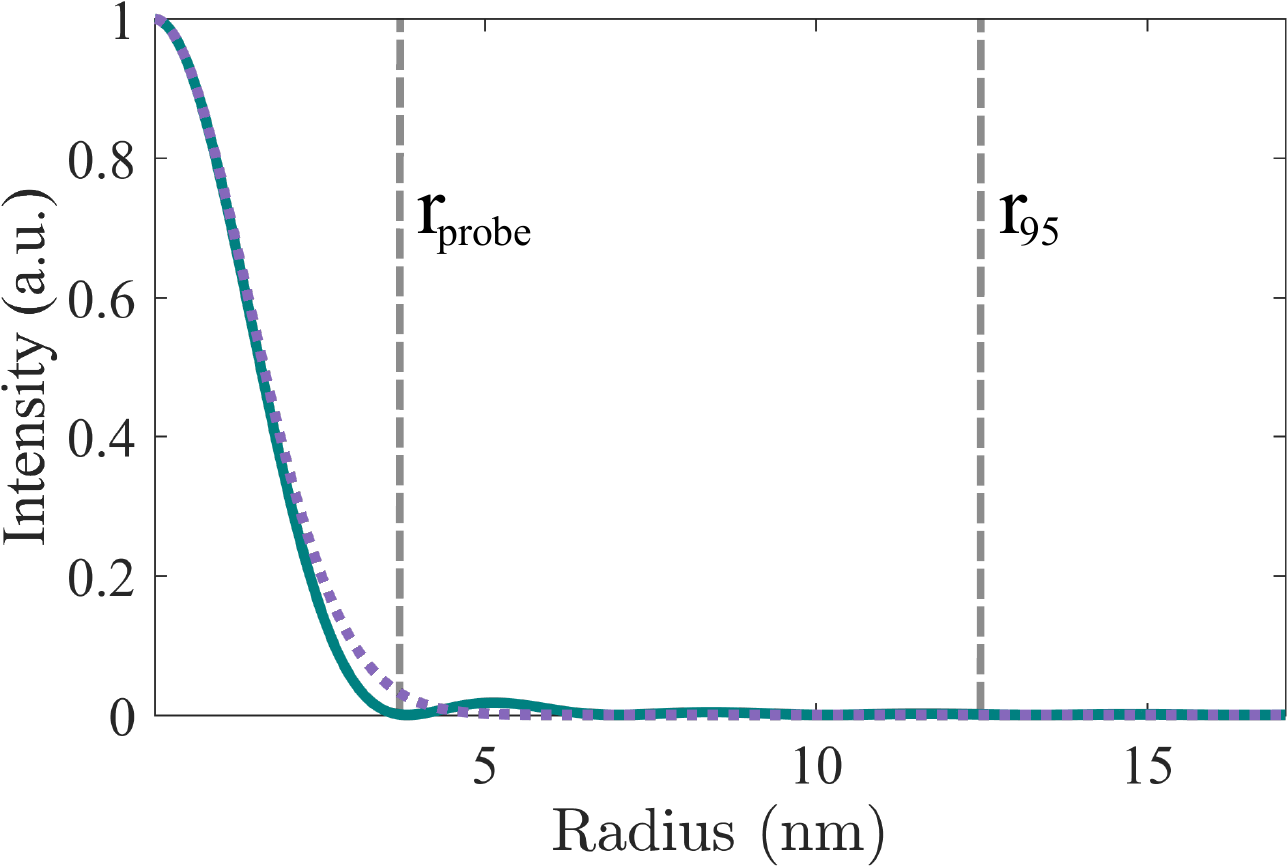}
\label{fig:GaussianAiryINTPlots}}\\
\subfloat[][]{
\includegraphics[width=0.9\columnwidth]{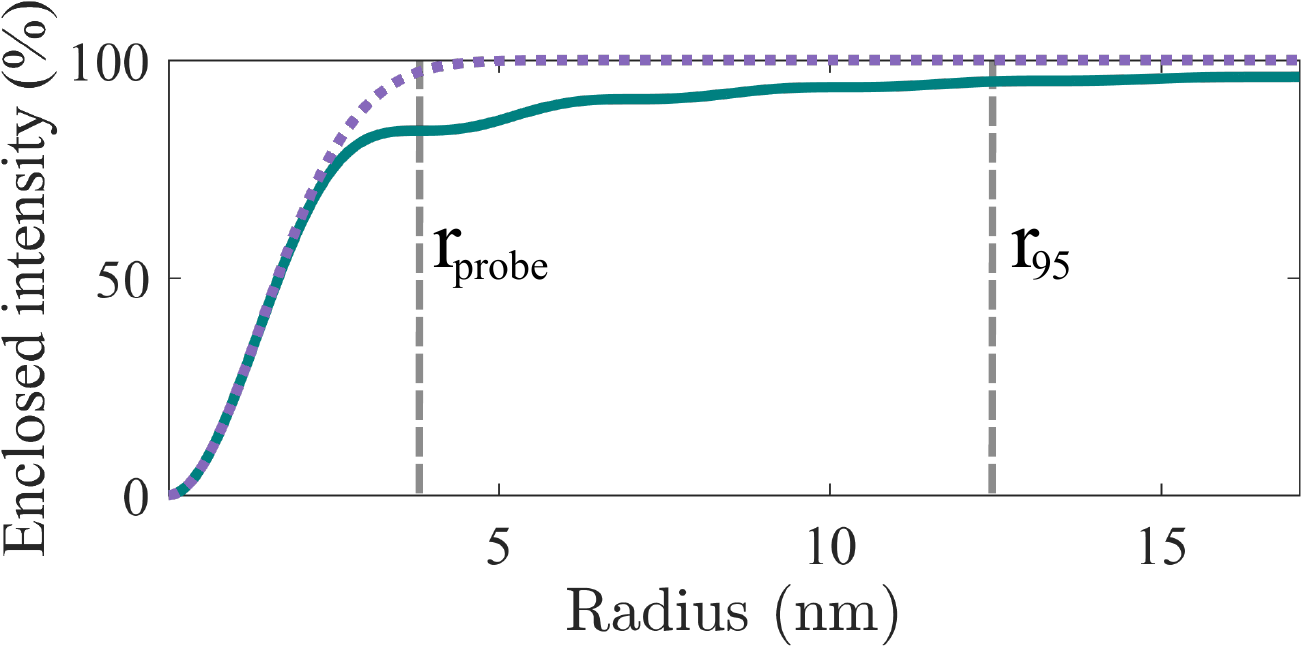}
\label{fig:GaussianAiryEncPlots}}
\caption{(a) Comparison of Airy (teal, solid) and Gaussian (lilac, dashed) intensity profiles, (b) Comparison of enclosed energy for the Airy and Gaussian probes. These plots assume $2 \pi k_0 \alpha = 1\, \mathrm{nm}^{-1}$; for suitable scaling via Eq. (\ref{eq:AiryIntensity}) these distributions can apply to any $k_0$ and $\alpha$ combination.}
\label{fig:PlotAiryGauss}
\end{figure}

The breakdown of the rigid-intensity-shift model is exacerbated by the long tails of the STEM probe intensity.  In the absence of lens aberrations the entrance wavefunction in STEM is given by:
\begin{equation}
|\Psi_{\rm entrance}(x,y)|^2 = I_0 \left[ \frac{J_1 (2 \pi k_0 \alpha r)}{2 \pi  k_0 \alpha r} \right]^2,
\label{eq:AiryIntensity}
\end{equation}
where $I_0$ is an intensity normalisation, $r=\sqrt{x^2+y^2}$, $J_1$ is a first-order Bessel function of the first kind, and $k_0=1/ \lambda$.  Such a probe is well known, comprising a central Airy disc, and surrounded by weaker rings with intensities decreasing slowly with radius.  This is shown as an intensity profile in Fig.~\ref{fig:PlotAiryGauss}a.  In STEM imaging, the probe size is typically referred to as $r_{\rm probe}=0.61\  \lambda / \alpha$, the radius of the central disc.  However, as shown in the enclosed intensity plotted in Fig.~\ref{fig:PlotAiryGauss}b, this central disc contains only $84$\% of the probe intensity.  The remaining intensity is broadly distributed, as demonstrated by comparison with a Gaussian probe of the same full-width at half-maximum (FWHM) \cite{bornwolf, mckechniegeneral}.  If instead we define the probe radius as $r_{95}=1.96 \ \lambda / \alpha$, the radius which contains $95$\% of the probe intensity, the potential for inaccurate data interpretation in DPC-STEM becomes clear: the probe has a much larger spatial extent in real space than is typically accounted for.

For atomic resolution STEM imaging, the gradient of the phase profile imparted by individual atomic columns varies significantly across the incident wavefunction of even the narrowest aberration-corrected STEM probes, where the resultant elaborate diffraction pattern intensity redistribution is widely appreciated.
Analysis methods that are suited to this regime and exploit recent improvements in high-speed pixellated detectors \cite{tate2016high,yang2016simultaneous} include 
 first-moment-detector DPC-STEM \cite{waddell1979linear,muller2014atomic,lubkzweck,lazic2016phase,muller2017measurement,lazic2017chapter,cao2017theory}and ptychography \cite{Faulkner_2004_R,Morgan_2013_DWSKA,DAlfonso_2014_MYWSKA,yang2016simultaneous}.  While these new detectors seem promising, at present they remain specialist systems which produce extremely large datasets \cite{lazic2017chapter} and do not readily allow live imaging (i.e. at the few-second refresh rate of experimental scans) to find structural features of interest and optimise the probe aberrations.  One compromise is a segmented detector system, as depicted in Fig.~\ref{fig:DPCschematic}, which uses more established technology, allows for faster scans and produces more manageably-sized data sets.  For strong phase objects, segmented detectors can give a very good approximation to the diffraction pattern centre-of-mass \cite{lazic2016phase,close2015towards,lazic2017chapter}.  For weak phase objects, they allow for linear reconstructions \cite{Landauer_1995_MR,MLR1,majert2015high,pennycook2015efficient,seki2017quantitative}.

While these more elaborate segmented detector analysis methods can be applied to long range fields \cite{brownmeasuring}, the rigid-intensity-shift model has much to recommend it in addition to its conceptual simplicity.  Zweck \emph{et al.} used its analytical tractability to establish clear guidelines for achievable field sensitivity \cite{zweckdetector,schwarzhuber2017achievable}.  Experimental calibration \cite{SFSMSKOMI1,schwarzhuber2017achievable,wu2017correlative} enables simple implementation and analysis, and can largely account for inelastic scattering effects from thicker samples \cite{brownmeasuring}.
In this paper we seek to better understand the diffraction pattern intensity redistribution in objects with long-range fields, to establish the domain of validity of the rigid-intensity-shift model of DPC-STEM, and to explore the manner in which it breaks down \footnote{Chapman \emph{et al.} \cite{chapman1978direct} explored this question in the context of domain wall imaging by Taylor expanding the imparted phase of Eq.~(\ref{eq:a1}) to second order.  Appropriate to the instrumentation of the time, the first order correction to the rigid-shift model depended on lens aberrations but vanishes in an aberration-free system.  The current generation of aberration corrected STEM instruments offer greater control over lens aberrations, and a particular strength of DPC-STEM is that it can be done in-focus, usually the optimum imaging conditions for other STEM imaging modes (such as high-angle annular dark-field \cite{williamscarter}) that one might wish to acquire simultaneously \cite{SFSMSKOMI1,shibatadirect}.}.

This paper is organised as follows.  In Sec.~\ref{pnsection}, a p-n junction in a Gallium Arsenide (GaAs) specimen is explored as a case study of a phase profile varying in one dimension (i.e. $\nabla \phi$ is a function of $x$ alone), using both experimental data and an analytical model to investigate realistic limits to the rigid-disk-shift model of DPC-STEM.  A simulation study of magnetic diamond domains in a Nickel-Iron (NiFe) specimen is explored as a case study of a 2D-varying phase profile in Sec.~ \ref{sssec:DDD}.  These case studies demonstrate that while the rigid-intensity-shift model does not exactly hold, for imaging long range fields the more complex intensity redistribution occurs predominantly near the edges of the diffraction pattern.  As such, Sec.~ \ref{sec:segdet} explores the precision obtainable for quantitative field imaging with a segmented detector when analysed in the rigid-intensity-shift model.  Since the extent of STEM probe tails is shown to be a limiting factor, Sec.~ \ref{suggestion} proposes a beam shaping strategy to extend the validity of the rigid-intensity-shift model.

\section{1D-varying phase profile case study: \texorpdfstring{$\mathbf{p-n}$}{p-n} junction in \texorpdfstring{$\mathbf{GaAs}$}{GaAs}} \label{pnsection}

Here we revisit and extend the previously-examined case of a 290 nm thick, slab-like specimen of GaAs with symmetrical p-n junction between $10^{19}$ cm$^{-3}$ p-doped (Zn) and $10^{19}$ cm$^{-3}$ n-doped (Si) regions  \cite{SFSMSKOMI1,brownmeasuring}. 
For this system, the transmission function describing the phase profile imparted by the intrinsic electric field across the junction is well-approximated by the 1D function \cite{SFSMSKOMI1}:
\begin{equation}
T_{p-n}(x) = \exp\left[i\sigma t V_0 \mathrm{erf}\left(\frac{x}{d}\right)\right],
\label{eq:pnphase}
\end{equation}
where $\sigma$ is the interaction constant (7.29$\times 10^{-3}$~(V$\cdot$nm)$^{-1}$ for 200~keV electrons) \cite{degraefbook}, $t=67$ nm is the (deduced) active-region thickness, $V_0=1.8$~eV is the difference in mean inner potential between the p- and n-doped regions of the semiconductor material, and $d=17$~nm is the characteristic width of the junction (numerical values as determined in Ref. \cite{SFSMSKOMI1}).

\begin{figure}[htb!]
\centering
\subfloat[][]{
\includegraphics[width=0.8\columnwidth]{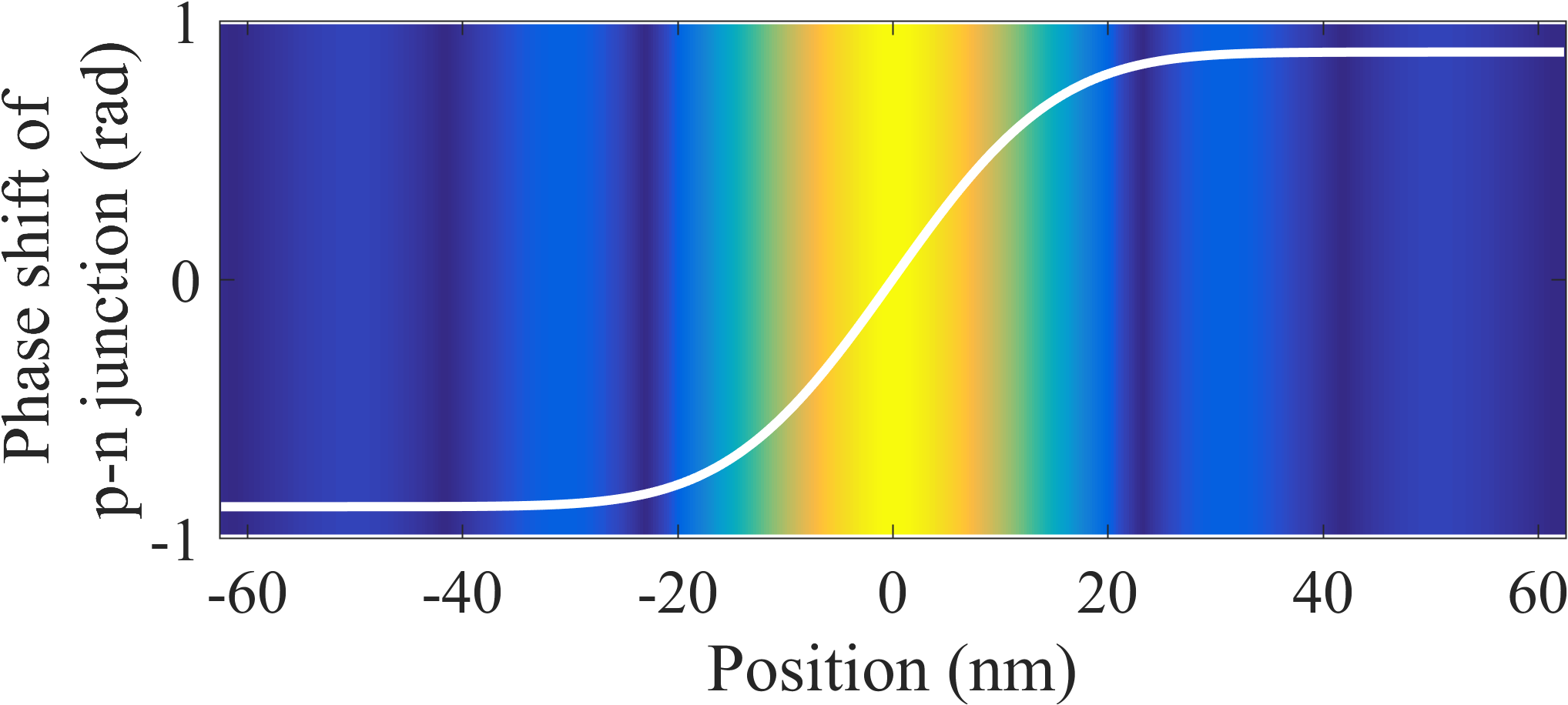}
\label{fig:subfhgetrsig2}}\\
\subfloat[][]{
\includegraphics[width=0.8\columnwidth]{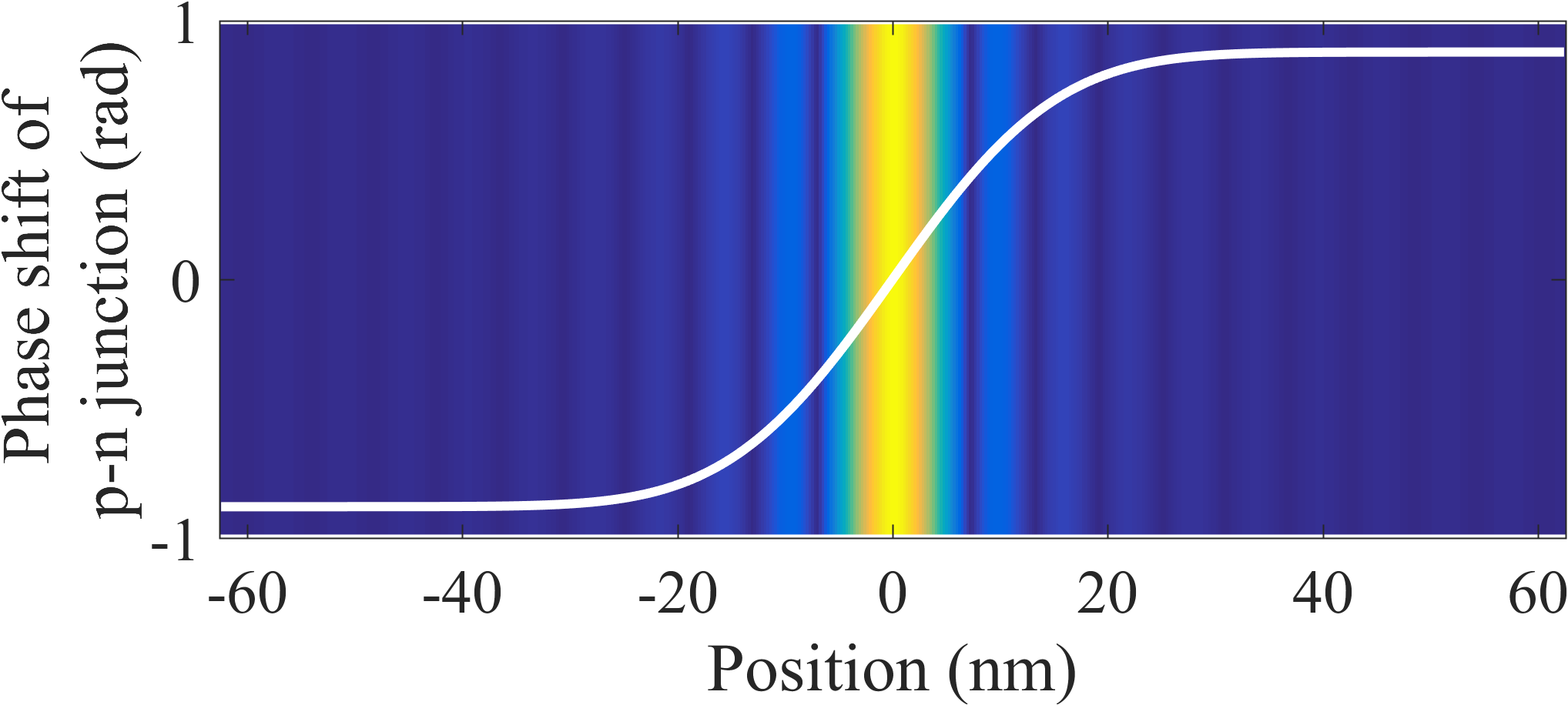}
\label{fig:subfihtgrdjg2}}\\
\subfloat[][]{
\includegraphics[width=0.8\columnwidth]{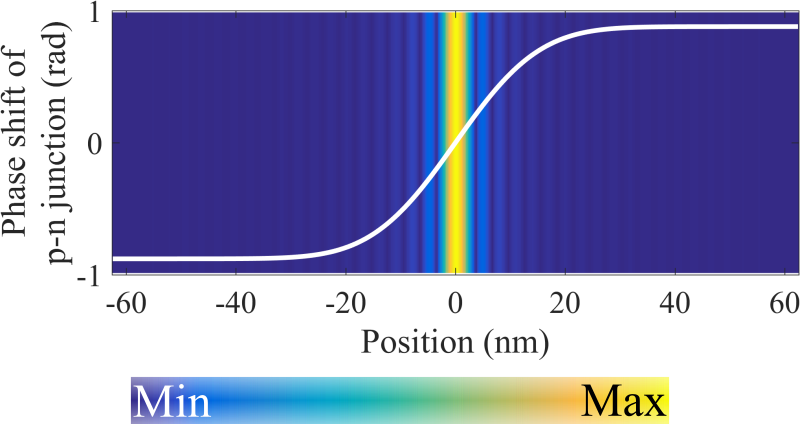}
\label{fig:jtyrj}}\\
\subfloat[][]{
\includegraphics[width=0.8\columnwidth]{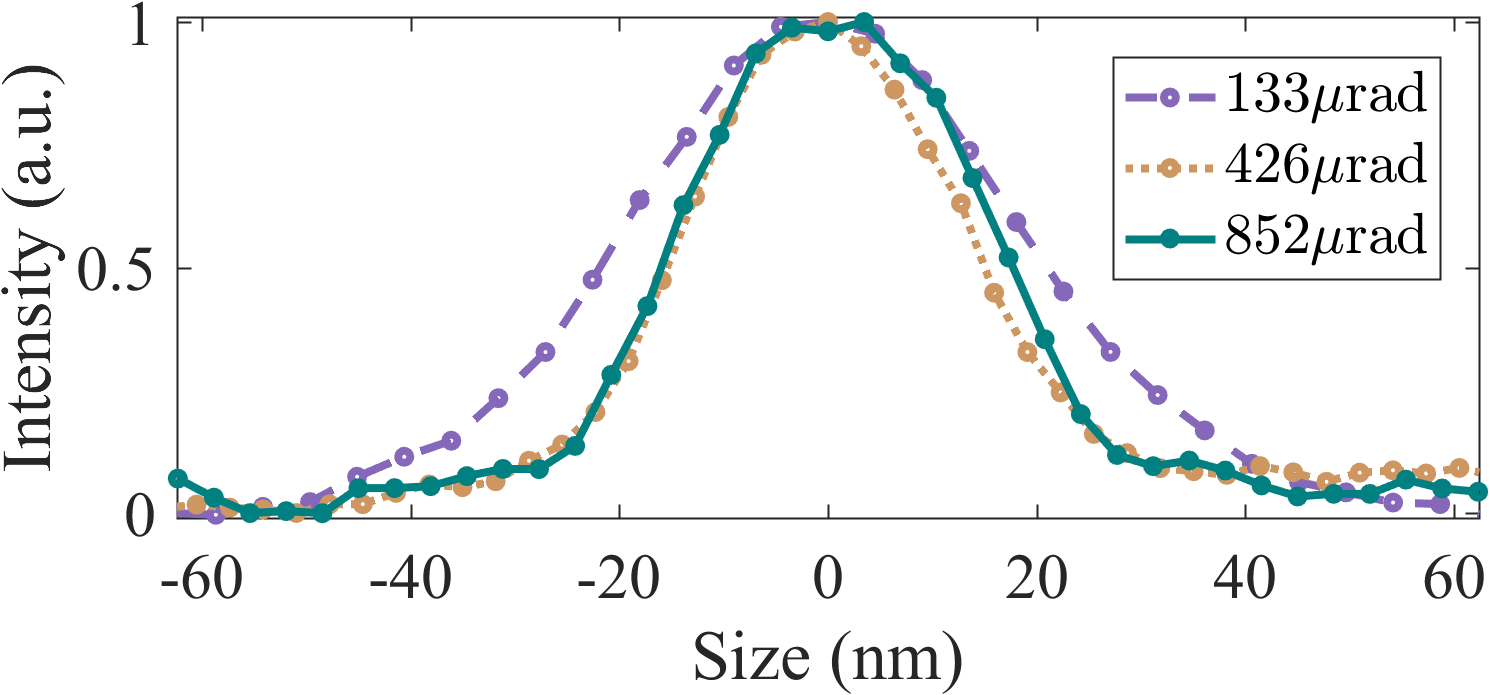}
\label{fig:subfigrdg1}}
\caption{Plots of the p-n junction phase profile, underlaid by a colormap of the probe amplitude profile, extended vertically to aid visual comparison between the variation in curvature of the junction phase profile and spatial extent of the probe amplitude profile, for the probe-forming aperture semiangles (a) $\alpha=133\ \mu$rad, (b) $\alpha= 426\ \mu$rad, and (c) $\alpha=852\ \mu$rad. (d) Experimental p-n junction DPC-STEM profiles as imaged with the three different probes.  These profiles were obtained by taking the difference between the STEM images from the two diametrically-opposed detector segments under the edges of the bright field disk and integrating the result along the length of the junction.}
\label{fig:DPCresolution}
\end{figure}

In seeking to understand the limits of the rigid-disk-shift model, we illuminated this specimen with three different probe sizes, characterised by convergence semiangles $\alpha=$ 133, 426 and 852 $\mu$rad, the scaling of which are shown in comparison to that of the phase profile of the transmission function of the junction in Figs.~\ref{fig:subfhgetrsig2}, \ref{fig:subfihtgrdjg2}, and \ref{fig:jtyrj}, respectively \footnote{In contrast to Fig.~\ref{fig:PlotAiryGauss}, the probe amplitude, $|\Psi_{\rm entrance}|$, is shown here to emphasise the extent of the probe tails.}.  The $\alpha = 133 \; \mu$rad case produces the broadest probe, and is that used previously \cite{SFSMSKOMI1,brownmeasuring} for which the diffraction pattern showed an intensity redistribution more complex than a simple rigid intensity shift.  This followed because the widths of the p-n junction and probe intensity distribution are comparable: the phase gradient varies appreciably across the central intensity lobe of the probe distribution, as seen in Fig.~\ref{fig:subfhgetrsig2}.  We might therefore expect that the finer probes, for which Figs.~\ref{fig:subfihtgrdjg2} and \ref{fig:jtyrj} show less variation in the phase gradient across the region of appreciable intensity, would better justify the rigid-disk-shift model.  This expectation is reinforced by the DPC-STEM profiles in Fig.~\ref{fig:subfigrdg1} which converge to essentially the same profile for the two narrower probes.

\begin{figure*}[htb!]
\includegraphics[width=0.9\textwidth]{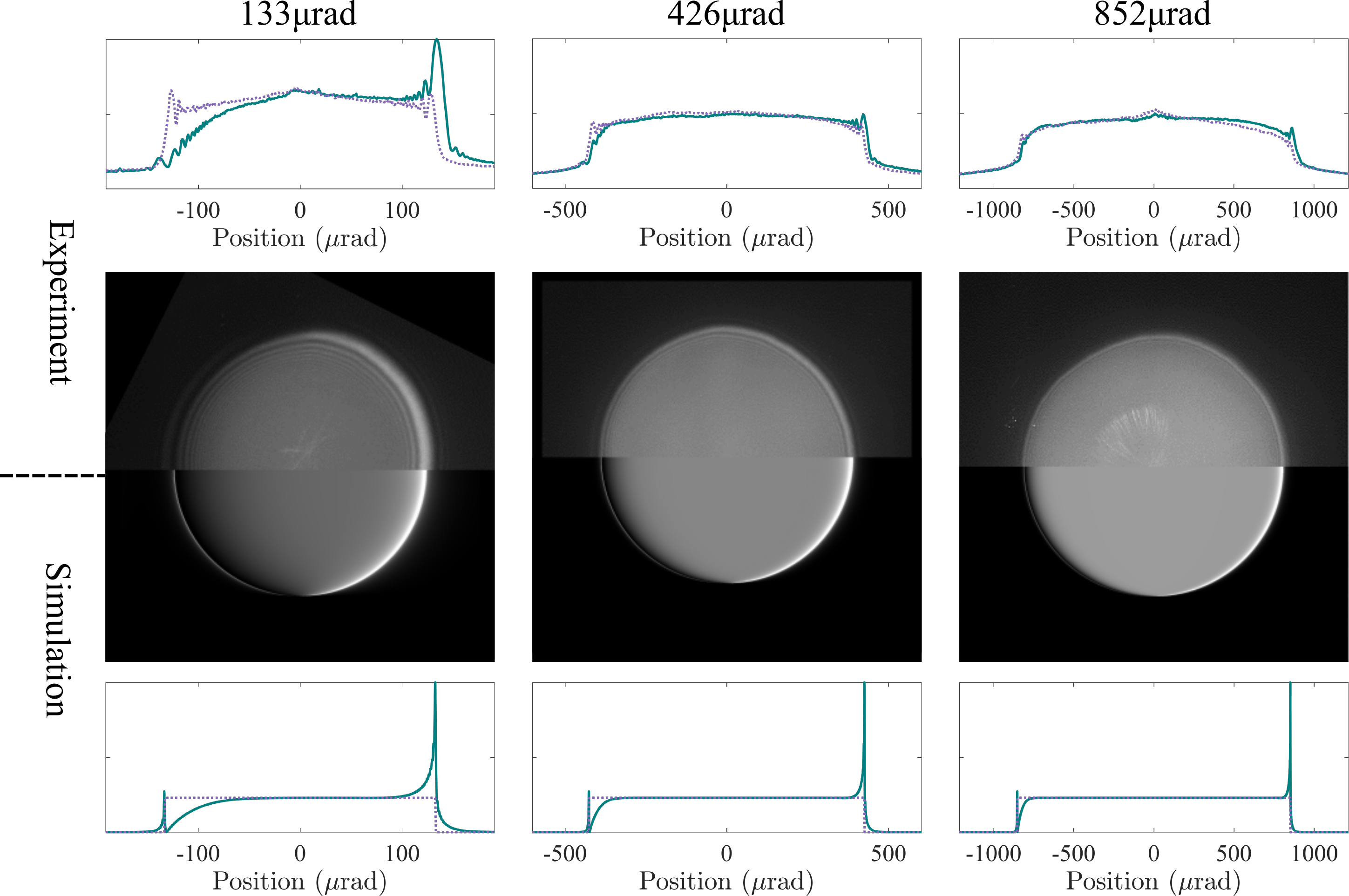}
	\caption{Line profiles and 2D diffraction patterns comparing experiment and simulation for three different convergence semiangles, illuminating the p-n junction specimen.  The line profiles compare results on-junction (teal, solid) with those off-junction (lilac, dashed). \label{fig:blerghhh}}
\end{figure*}

The detailed diffraction pattern distributions, however, show that the scattering physics is not so simple.  Figure~\ref{fig:blerghhh} compares diffraction patterns between experiment and simulation (using Eq.~(\ref{eq:ronchiInt})) for the three different convergence semiangles.  The intensity profiles, taken from across the centre of the full diffraction patterns, compare on-junction to off-junction results.  The experimental and simulated patterns are in broad qualitative agreement. (The Fresnel fringing and other fine structure evident in the experimental patterns result from the images having been recorded on photographic film and so containing residual aberrations that were not able to be identified and minimised during recording.)
A rigid-disk-shift model would predict a shift of approximately 18 $\mu$rad (based on the field strength at the centre of the junction), but the patterns make clear that none of the on-junction patterns are simply rigidly shifted versions of their off-junction counterparts.  Rather, each pattern shows a bright-intensity peak on the right hand edge of the disk, and a reduced-intensity trough on the left hand edge of the disk, at diffraction-plane positions broadly within the same area as that illuminated in the off-junction case.  Indeed, although these peak and trough features constitute a smaller fraction of the diffraction pattern for increasing convergence semiangle, their angular extent is the same in each case.

\begin{figure*}[htb!]
	\subfloat[]{\includegraphics[width=0.33\textwidth]{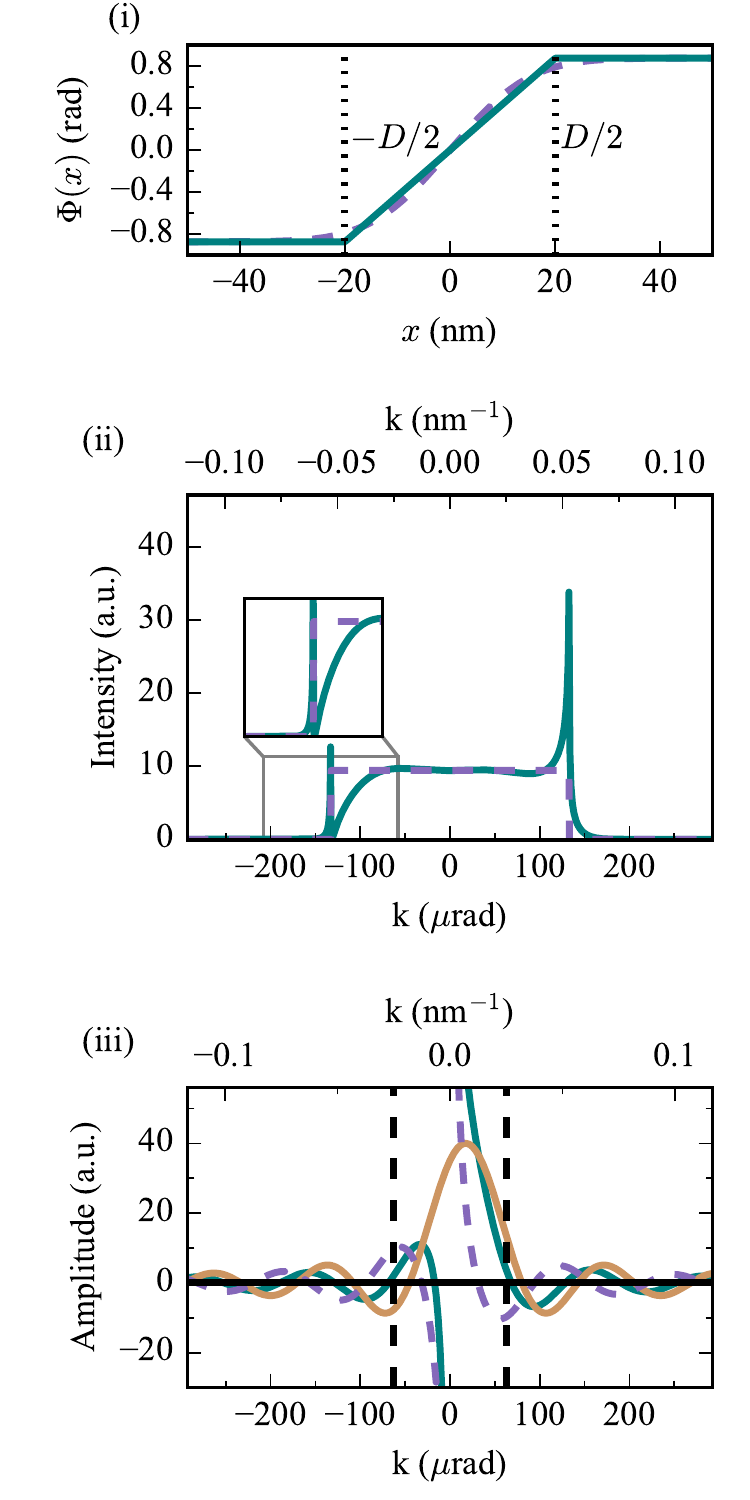}}
	\subfloat[]{\includegraphics[width=0.33\textwidth]{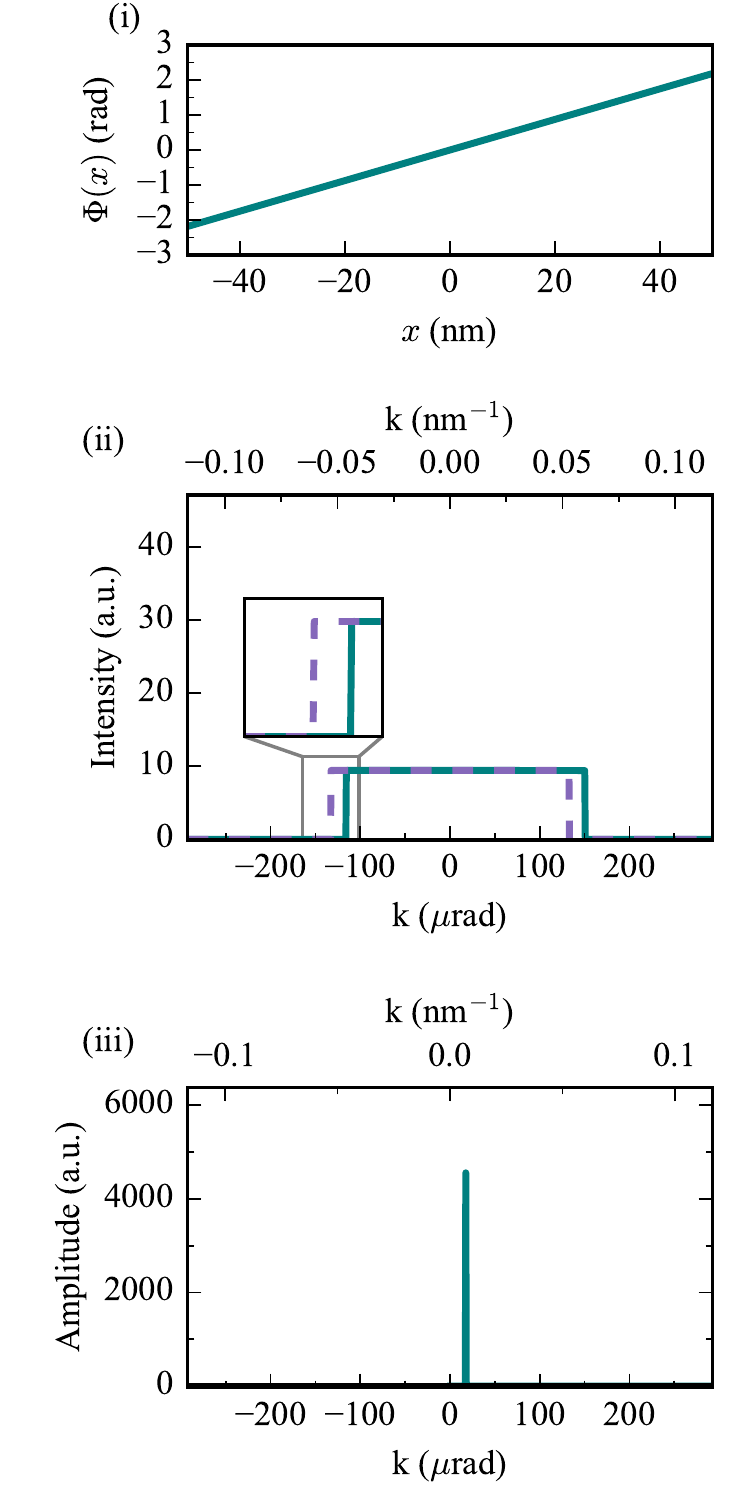}}
	\subfloat[]{\includegraphics[width=0.33\textwidth]{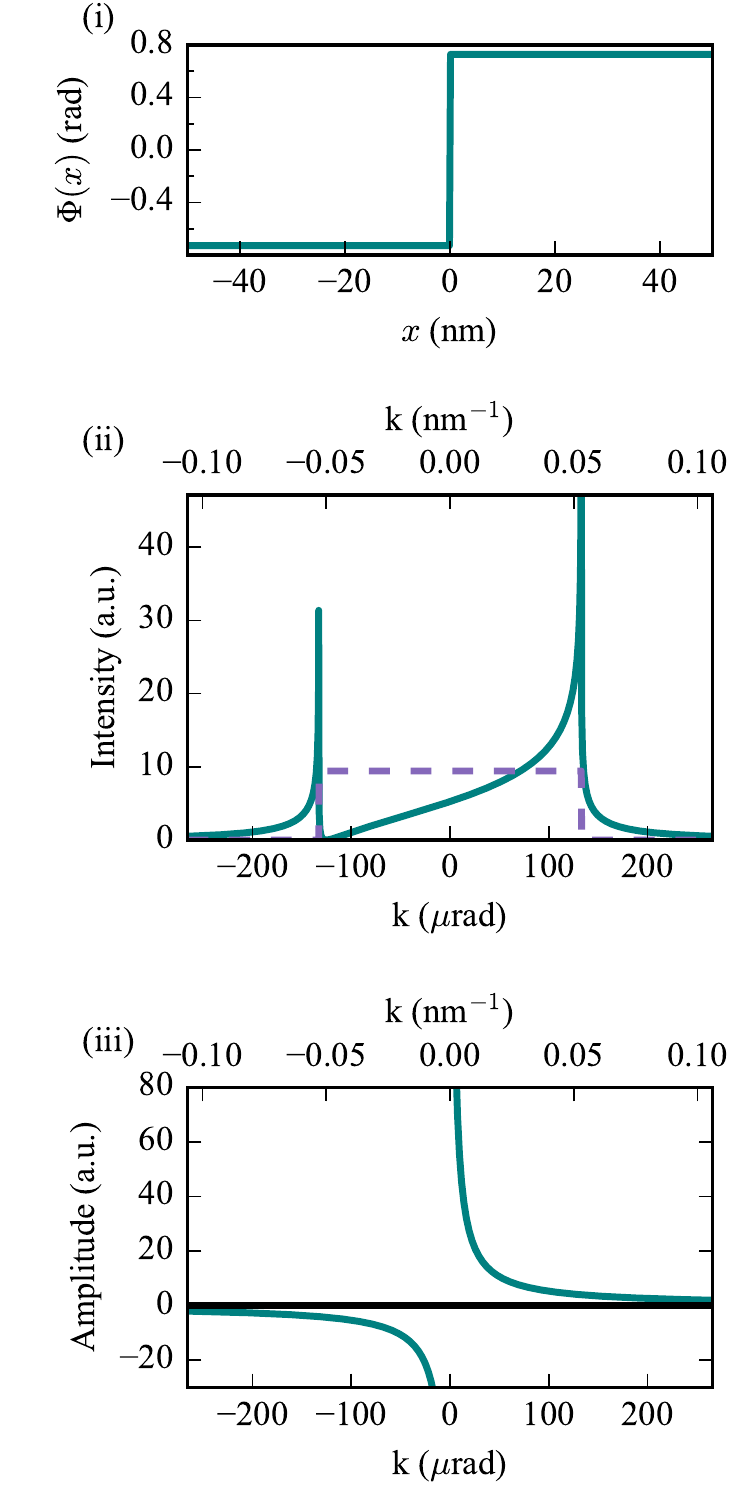}}
	\caption{For (a) the piecewise p-n junction model of Eq. (\ref{eq:constantE}), (b) the $D\rightarrow \infty$ limit and (c) the $D\rightarrow 0$ limit we plot the (i) real-space transmission function phase, (ii) diffraction pattern, and (iii) reciprocal-space transmission function. In (a)(i) piecewise p-n junction model phase (teal, solid), is compared to the continuous equivalent function (lilac, dashed).  In parts (ii), the diffraction pattern on-junction (teal, solid) is compared to that off-junction (lilac, dashed).  In (a)(iii) the reciprocal-space transmission function (teal, solid) is compared to the $2^{nd}$ term of Eq.~(\ref{eq:9}) (lilac, dashed) and the $3^{rd}$ term of Eq.~(\ref{eq:9}) (mustard, solid), with vertical black lines positioned at $\pm 1/D$ as guides to the eye.\label{fig:analyticmodelmegaplot}}
\end{figure*}

To better understand these features, let us consider a piecewise approximation to Eq.~(\ref{eq:pnphase}) that is amenable to analytic manipulation.  Assuming a constant electric field within the p-n junction and zero electric field outside, the transmission function may be written:
\begin{align}
T(x) = 
\begin{cases}
e^{-i\varphi D/2} & \; x<-D/2\\
e^{i\varphi x} & |x| \leq D/2\\
e^{i\varphi D/2} & \; x>D/2\\
\end{cases}\,,
\label{eq:constantE}
\end{align}
where $D$ is the nominal width of the junction and $\varphi =\sigma V_0t/D$. Note that $D$ is different to the characteristic width, $d$, in the error function model of Eq.~(\ref{eq:pnphase}).
For $d = 17$~nm, a value of $D = 46$~nm minimizes the root-mean-square error between this piecewise approximation and the error function model, the comparison is shown in Fig.~\ref{fig:analyticmodelmegaplot}a(i). 

Equation~(\ref{eq:constantE}) can be rewritten in terms of functions commonly found in tables of Fourier transforms:
\begin{align}\nonumber
T(x) =& \cos(\varphi D/2)[1-\mathrm{Rect}_D(x)]\\ \nonumber
&+i\frac{\sin(\varphi D/2)}{2}\left[\delta(x-D/2)+\delta(x+D/2)\right]\otimes\mathrm{sgn}(x)\\
&+e^{i\varphi x}\mathrm{Rect}_D(x),
\label{eq:8}
\end{align}
where $\mathrm{sgn}(\pm|x|)=\pm1$ is the sign function and $\mathrm{Rect}_D(x)$ is the rectangle function given by:
\begin{align}
\mathrm{Rect}_D(x) = \begin{cases}
1 & |x|<D/2\\
0 & \text{otherwise}
\end{cases}\,.
\end{align}
The first two terms in Eq.~(\ref{eq:8}) are only non-zero for $|x| \ge D/2$ and so pertain to the field-free region of the specimen.  The final term is only non-zero for $|x|<D/2$ and so pertains to the region of the specimen where the electric field is constant.  Fourier transformation of Eq.~(\ref{eq:8}) gives:
\begin{align}\nonumber
T(k) =& \cos(\varphi D/2)[\delta(k)-D \, \mathrm{sinc}(\pi D k)]\\
&+\frac{\sin(\varphi D/2)\cos(\pi D k)}{\pi k}+D \, \mathrm{sinc}(\pi D (k-\varphi/2\pi))\,.
\label{eq:9}
\end{align}
With reference to Eq.~(\ref{eq:ronchiInt}), the diffraction plane wavefunction of the scattered probe is given by the convolution of Eq.~(\ref{eq:9}) with the reciprocal space illumination wavefunction (the aperture function). In 1D, the aperture function is a top-hat and, for comparison with the experiments,  we have set the width to be $\alpha = 133$ $\mu$rad. The resulting diffraction pattern is shown in Fig.~\ref{fig:analyticmodelmegaplot}a(ii) as a teal, solid line. For reference, the reciprocal-space form of the entrance wavefunction intensity $|\Psi_{\mathrm{entrance}}|^2$, the aperture function, is also shown as a lilac, dotted line.  It can be seen that the simplified analytic model qualitatively accounts for the features of the diffraction pattern plotted in Fig.~\ref{fig:blerghhh} for the $\alpha=133 \; \mu$rad case, which used the error function model for the phase of the p-n junction transmission function.  Different components of Eq.~(\ref{eq:9}) are plotted in Fig.~\ref{fig:analyticmodelmegaplot}a(iii). Before discussing these in detail, it is helpful to consider two limiting cases.

The third term in Eq.~(\ref{eq:9}), $D\,\mathrm{sinc}[\pi D (k-\varphi /2\pi)] $, corresponds to the region of constant electric field in Eq.~(\ref{eq:constantE}) and accounts for a shift in the diffraction pattern centre of mass due to the transverse electric field of the specimen. In the limit $D\gg 1$, this $\mathrm{sinc}$ term will approach a $\delta$-function (i.e. cause a rigid shift of the illumination).  Figure~\ref{fig:analyticmodelmegaplot}b(i) plots the transmission function phase assuming the limiting case of a very large p-n junction with the same built-in electric field ($|\nabla \phi| = 3.68\times 10^{-2}$ rad/nm).  This causes the diffraction pattern to shift rigidly to the right (by $5.83\times 10^{-3}$ nm$^{-1}$ or 14.6 $\mu$rad), as shown in Fig.~\ref{fig:analyticmodelmegaplot}b(ii).  The Fourier transform of the transmission function is seen to essentially be a $\delta$-function, Fig.~\ref{fig:analyticmodelmegaplot}b(iii).

In the limit $D\rightarrow 0$, but with $\varphi$ scaled such that $\varphi D$ is constant, Eq.~(\ref{eq:9}) approaches:
\begin{align}
T(k) =& \cos(\varphi D/2)\delta(k)+\frac{\sin(\varphi D/2)}{\pi k}\,.
\label{eq:10}
\end{align}
Here it is possible to derive an analytic expression for the diffraction pattern:
\begin{align}
|\Psi(k)|^2= &\frac{1}{2\alpha  k_0}\Bigg[\cos(\varphi D/2)\mathrm{Rect}_{\alpha k_0}(k) \nonumber \\
&+\frac{\sin(\varphi D/2)}{\pi}\ln\left|\frac{k+\alpha k_0}{k-\alpha k_0}\right|\Bigg]^2 \;.
\label{eq:12}
\end{align}
Setting $\varphi D = 1.76$ rad, to produce the same potential difference as that across the p-n junction in Fig.~\ref{fig:analyticmodelmegaplot}a(i), gives the step-function transmission function phase shown in Fig.~\ref{fig:analyticmodelmegaplot}c(i).  The  diffraction pattern resulting from Eq.~(\ref{eq:12}) is plotted in Fig.~\ref{fig:analyticmodelmegaplot}c(ii).  The most pronounced features in this diffraction pattern are the sharp peaks at $k=\pm \alpha k_0$, i.e. at the edges of the aperture function, resulting from the logarithmic divergence in Eq. (\ref{eq:12}).  Note that $\alpha$ is the only meaningful length scale in this limit, and as such the intensity both within and spreading beyond the aperture function varies on this scale.  Figure~\ref{fig:analyticmodelmegaplot}c(iii) plots the transmission function $T(k)$, showing the divergence inherent in the $1/k$ factor in the second term of Eq. (\ref{eq:10}).  This establishes why the points of divergence in the diffraction pattern occur at the edges of the aperture function: in convolving the top-hat function with this transmission function, those are the points where the top-hat overlaps only one half of the $k \rightarrow 0$ divergence.

Figure~\ref{fig:analyticmodelmegaplot}b and the associated discussion showed how the $\mathrm{sinc}$ term in Eq.~(\ref{eq:9}) gives rigid-intensity-shift behaviour.
Similarly, Fig. \ref{fig:analyticmodelmegaplot}c and the associated discussion showed how the $1/k$ term in Eq.~(\ref{eq:9}) leads to the sharp peaks at the edges of the aperture function. 
These observations aid interpretation of the relative contribution of the different terms in Eq.~(\ref{eq:9}) to the diffraction pattern shown in Fig.~\ref{fig:analyticmodelmegaplot}a(ii) resulting from the piecewise approximate \mbox{p-n} junction potential.  Figure~\ref{fig:analyticmodelmegaplot}a(iii) explores the relative contributions of the second and third term in Eq.~(\ref{eq:9}).  The transmission function $T(k)$ is plotted as a teal, solid line; the second term, $\sin(\varphi D/2)\cos(\pi D k)/\pi k$, as a lilac, dashed line; and the third term, the $\mathrm{sinc}$ term, as the mustard, solid line.

It can be seen that $T(k)\rightarrow \pm \infty$ as $k\rightarrow 0$ due to the $1/k$ factor in the second term in Eq.~(\ref{eq:9}), which necessarily dominates for sufficiently small $k$ and, as seen in discussion of Fig. \ref{fig:analyticmodelmegaplot}c, leads to sharp intensity peaks at the edges of the aperture function in Fig.~\ref{fig:analyticmodelmegaplot}a(ii), again the dominant feature of the diffraction pattern.  Note, however, that there is now an additional length scale in the problem: the junction width, $D$.  Through the $\cos(\pi D k)$ factor in the second term in Eq.~(\ref{eq:9}), this has the effect of making the intensity peaks at the edges of the aperture function in Fig.~\ref{fig:analyticmodelmegaplot}a(ii) narrower than those of Fig.~\ref{fig:analyticmodelmegaplot}c(ii).

While in the $D\rightarrow \infty$ limit in Fig.~\ref{fig:analyticmodelmegaplot}b(iii) the $\mathrm{sinc}$ term containing the rigid shift tendency was both $\delta$-function like and dominant, in Fig.~\ref{fig:analyticmodelmegaplot}a(iii) it has finite width (of order $1/D$) and the shift of the central peak is hidden within the divergence of the $\sin(\varphi D/2)\cos(\pi D k)/\pi k$ term.  The former means that the width of the intensity variation in the extended peak-trough feature is about $1/D=0.025$ nm$^{-1}$, or about 60 $\mu$rad, consistent with Fig.~\ref{fig:analyticmodelmegaplot}a(ii).  The latter means that the shift of the diffraction pattern intensity is obscured by the peak-trough feature.

\begin{figure}
\includegraphics[width=0.38\textwidth]{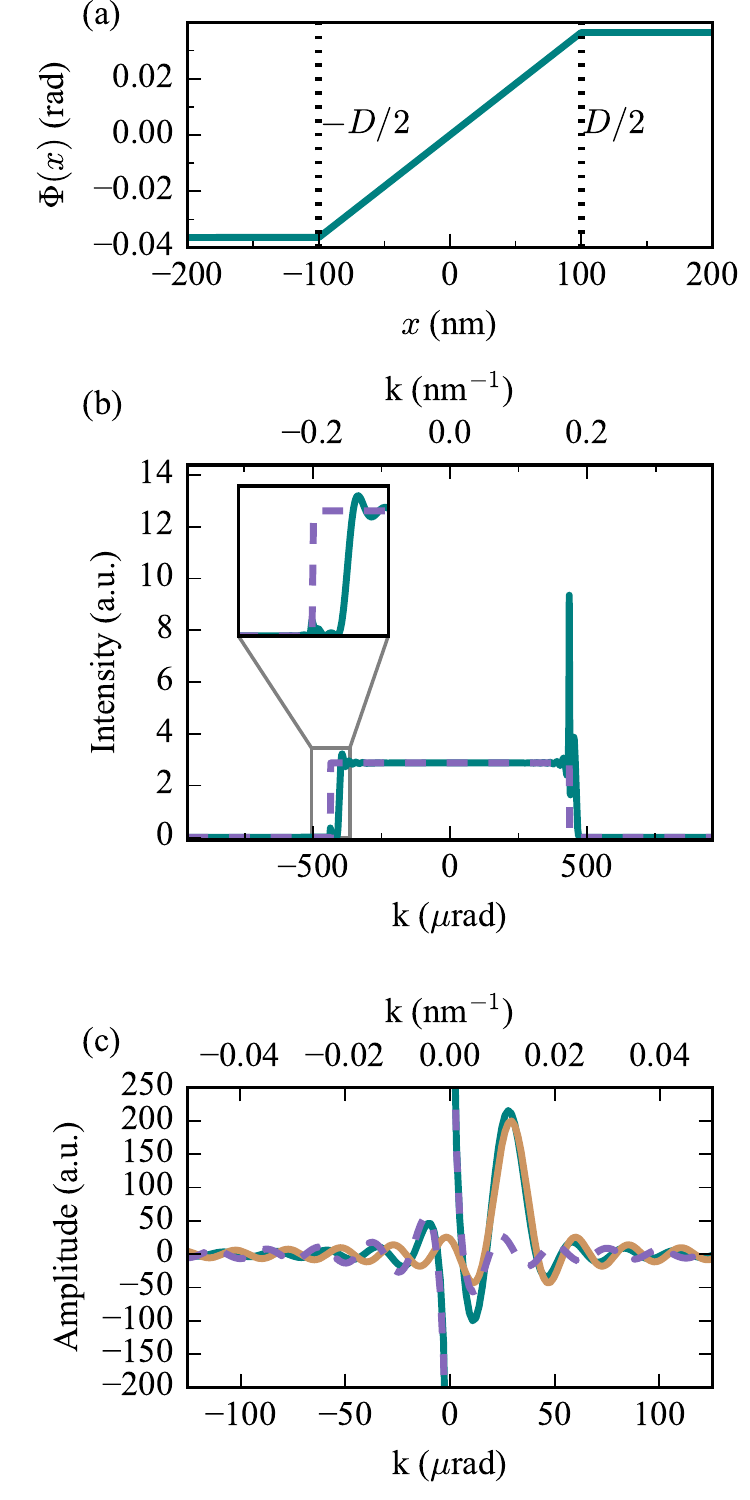}
	\caption{(a) Real-space transmission function phase, (b) diffraction pattern and (c) the reciprocal-space transmission function for the for the piecewise object of Eq.~(\ref{eq:constantE})  for parameters taken for a transmission function of a magnetic domain taken from Ref.~\cite{krajnakthesis}. The line styles are as per Fig.~\ref{fig:analyticmodelmegaplot}(a).	Note that (b) and (c) are plotted with different horizontal scales, because the features in (b) are an order of magnitude smaller than those in Fig.~\ref{fig:analyticmodelmegaplot}a(ii).
	\label{fig:krajnak}}
\end{figure}

It is also instructive to consider a case where the rigid-disk-shift model is known to be a good approximation. Krajnak shows model data from a polycrystalline magnetic sample which imposes a linear phase gradient of $\nabla \phi = 0.073$~rad/nm on the wavefunction in a region that extends for 200~nm, which is much larger than the probe ($r_{95}=11.3$ nm for a 436~$\mu$rad probe forming aperture at 200~keV)~\cite{krajnakthesis}.  Figure~\ref{fig:krajnak}a is the same as Fig.~\ref{fig:analyticmodelmegaplot}a(i) except that parameters pertinent to the Krajnak model have been used. As can be seen in Fig.~\ref{fig:krajnak}b, especially in the magnified inset, the intensity distribution in the diffraction pattern is rather well described by the rigid-disk-shift model, though on close inspection small intensity peaks at the edge of the aperture function are still evident.

We can understand this qualitative difference from Fig.~\ref{fig:krajnak}c, which explores the relative contributions of the second and third term in Eq.~(\ref{eq:9}).  Again, $T(k)\rightarrow \pm \infty$ as $k\rightarrow 0$ because the $1/k$ factor in the second term necessarily dominates for sufficiently small $k$, explaining the small intensity peaks at the edge of the aperture function.  Note, however, that these terms are smaller and narrower than the intensity peaks in Fig.~\ref{fig:analyticmodelmegaplot}a(ii) because the $\cos(\pi D k)$ factor in the second term oscillates rapidly (due to the large value of $D$) \footnote{These rapid oscillations might also be considered an example of the well-known Gibbs phenomenon \cite{goodmanintroduction} where if even a generous bandwidth limit is applied to a step discontinuity then high frequency oscillations will result near the discontinuity.  Here, the $\mathrm{Rect}_D(x)$ factor in the transmission function in Eq.~(\ref{eq:8}) acts as a bandwidth limit on the shifted component of the probe.}.  It is likely that such oscillations would be challenging to observe experimentally, even on state-of-the-art instruments, due to finite beam coherence and detector resolution, and further complicated by such oscillations being similar in form to Fresnel fringes resulting from imperfect focussing.  Also in contrast to Fig.~\ref{fig:analyticmodelmegaplot}a(iii), the $\mathrm{sinc}$ term containing the rigid-shift tendency is much narrower (because $1/D$ is smaller) and is clearly separated from the divergence point of the second term.  That the magnitude of the shift in $k$ is larger than the $1/D$ length scale on which the diffraction pattern intensity varies means that there is now a dominating shift of the diffraction pattern intensity.

The shift, $\varphi/2\pi = |\nabla \phi|/2\pi$, in the third term in Eq.~(\ref{eq:9}) will be greater than the $1/D$ length scale on which the diffraction pattern intensity varies if:
\begin{align}
\frac{|\nabla \phi|}{2\pi} \cdot D > 1\,.
\label{eq:13}
\end{align}
Noting that in the ideal rigid-disk-shift case the detected deflection angle, $\beta$, can be related to the imparted phase gradient (for small deflections) via \cite{krajnakthesis, zweckdetector}:
\begin{equation}
|{\nabla \phi}| = 2\pi k_0 \beta \;,
\label{eq:beetah}
\end{equation}
Equation~(\ref{eq:13}) can also be written as:
\begin{equation}
\beta > \frac{1/D}{k_0} \equiv \gamma
 \label{eq:gamma}
\end{equation}
where $\gamma$ is the scattering angle scale of the intensity redistribution at the edge of the diffraction patterns.
It is also interesting to note that recognising the mean momentum transfer to the probe as $\Delta p = h\Delta k \approx h|\nabla \phi|/2\pi=\hbar|\nabla \phi|$ and the size of the junction as $\Delta x=D$, then Eq.~(\ref{eq:13}) becomes analogous to the quantum mechanical Heisenberg uncertainty principle:
\begin{align}
\Delta p \Delta x > h\,.
\end{align}

By comparing the size of the different probes against the p-n junction in Fig.~\ref{fig:DPCresolution} we anticipated that, while the phase gradient varied appreciably on the scale of the $\alpha=133$ $\mu$rad probe, the much narrower $\alpha=852$ $\mu$rad probe would have shown a more rigid-intensity-shift-like behaviour.  However, this was not supported by the experimental and simulated results in Fig.~\ref{fig:blerghhh}.  The additional conditions of Eqs. (\ref{eq:13}) and (\ref{eq:gamma}) explain this: the deflection expected from the peak field strength in the p-n junction is $\beta \approx 18$ $\mu$rad, which is smaller than the length scale on which the diffraction pattern intensity varies, $\gamma \approx 60$ $\mu$rad.  This requirement is fundamental to the object but independent of the probe forming aperture, which is why forming a finer probe failed to make the scattering more rigid-shift like in Fig.~\ref{fig:blerghhh}.  Conversely, in the Krajnak example, $\beta \approx 29$ $\mu$rad is appreciably larger than $\gamma \approx 13$ $\mu$rad, hence the rigid-intensity-shift model holds better.

We now seek to broaden our understanding through a study of a more complex specimen geometry where $\nabla \phi$ varies in both $x$ and $y$.

\section{2D-varying phase profile case study: magnetic domains in \texorpdfstring{$\mathbf{NiFe}$}{NiFe}} \label{sssec:DDD}

DPC-STEM has been particularly useful for studying magnetic microstructure \cite{chapmanmodified, chapmaninvestigation,uhlig2004direct,uhlig2005shifting}. As such, phase gradients imposed by magnetic domains are a highly relevant model system.  Here we work with a simulated specimen of NiFe, with magnetisation vectors generated using the Object Oriented MicroMagnetic Framework (OOMMF) software developed at the National Institute of Standards and Technology (NIST) \cite{Donahue}. 
A standard soft magnetic material was modelled using anisotropy constant, $K=0$~J/m$^3$, saturation magnetization, $M_s=860$~kA/m, and exchange coefficient, $A=13$~pJ/m.

\begin{figure}[ht]
\centering
\subfloat[][]{
\includegraphics[width=0.8\columnwidth]{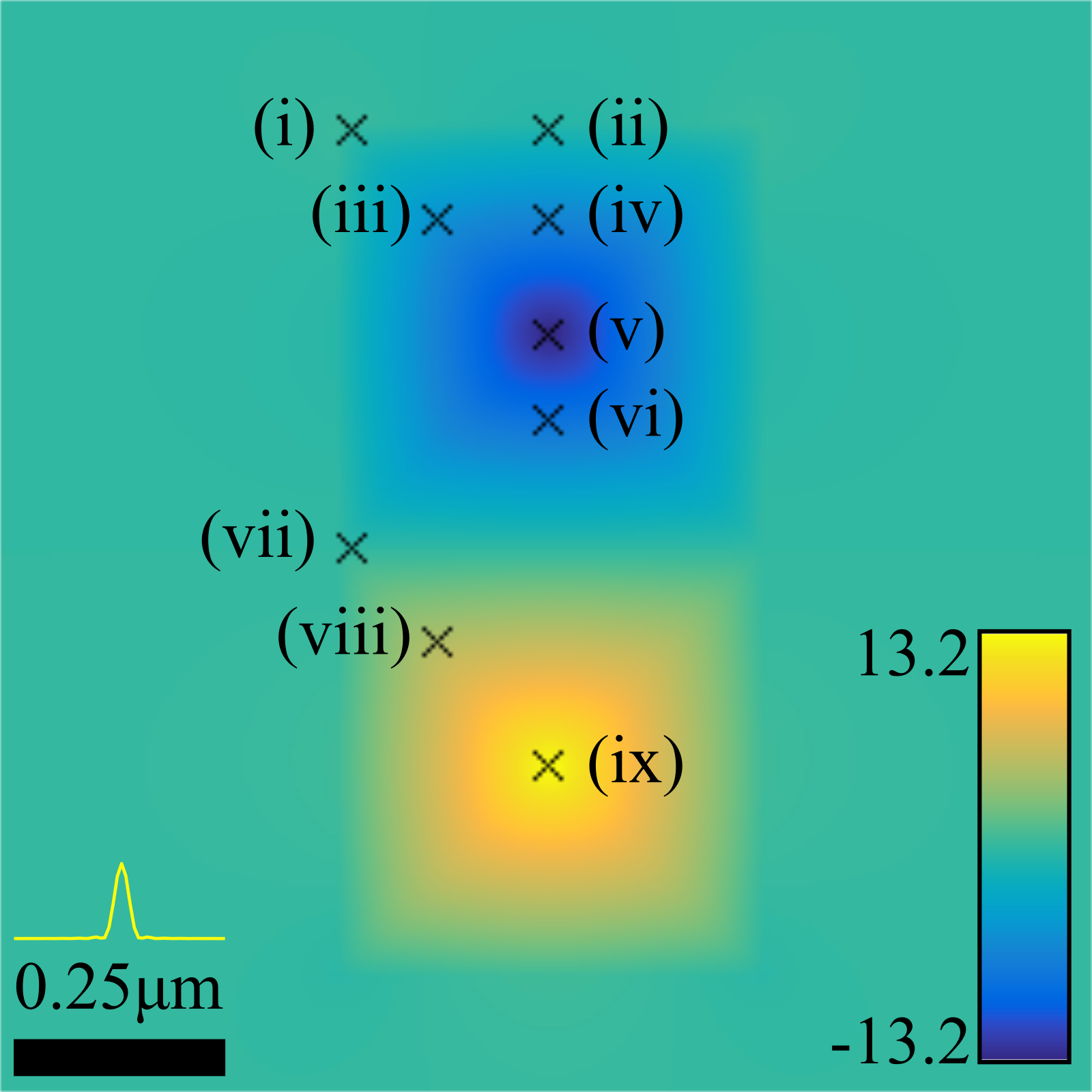}
\label{fig:labelledPhiAB}}\\
\subfloat[][]{
\includegraphics[width=0.8\columnwidth]{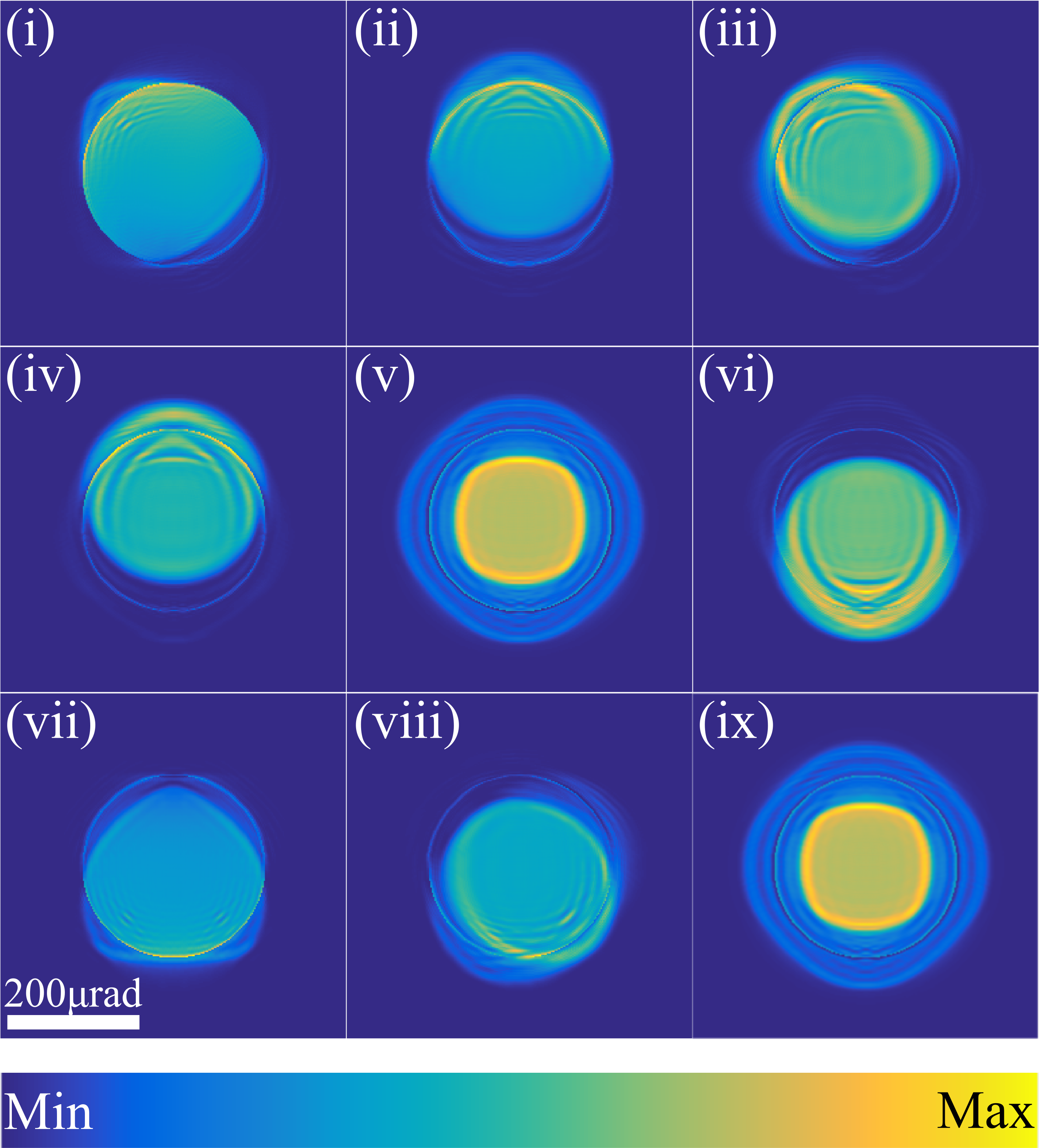}
\label{fig:DDDsuper2}}
\caption{(a) Phase profile imposed by the magnetised specimen ($1  \ \mu$m $\times 0.5 \ \mu$m), with colour bar in radians.  The crosses mark the probe positions for which the diffraction patterns in (b) are calculated, assuming a $133\, \mu$rad probe (the intensity profile of which is plotted above the scale bar in (a)). The intensity scale for each diffraction pattern has been set independently to best visualise the fine structure therein.}
\label{fig:DDDsuperr}
\end{figure}

The magnetisation vectors were converted to transmission function phase shifts following Refs. \cite{degraefphase, RDBchapter}, and interpolated to match the real-space requirements for adequately-sampled STEM diffraction calculations.  The electric potential is assumed constant.  The phase distribution imparted by this structure is illustrated in Fig.~\ref{fig:DDDsuperr}a.  The crosses indicate the probe positions at which the diffraction patterns shown in Fig.~\ref{fig:DDDsuperr}b were calculated, assuming a $133\ \mu$rad convergence semiangle.

In these diffraction patterns, a number of features are visible. Diffraction patterns (i) and (iii) both show the diagonal symmetry of the the phase profile at their respective probe positions, but with more severe peak-trough features in (iii) as more of the probe is sitting over regions of non-constant $\nabla \phi$. Diffraction patterns (ii), (iv) and (vii), with the least variation in phase gradient under the centre of the probe, show a general trend of intensity shift from their central position, but are still decorated with substantial intensity redistributions.  Fig.~\ref{fig:DDDsuperr}b(v) and (ix) are rather similar -- the probe positions have the same local symmetry though the phase profiles have opposite sign.  When $\nabla \phi$ varies dramatically under the central region of the probe, the intensity redistributions of the diffraction pattern become more severe.

The diffraction patterns in Fig.~\ref{fig:DDDsuperr}b show sharp peak-trough type features not unlike those seen earlier in Fig.~\ref{fig:blerghhh}: such features are not exclusive to the simple 1D-varying phase profile case but rather occur over a variety of systems and probe positions.  The generality of these features means that the rigid-intensity-shift model will rarely be exactly realised.  If the diffraction patterns in Fig.~\ref{fig:DDDsuperr}b were recorded on a pixel detector, the deviation from the rigid-intensity-shift model might itself be used to extract information about the structure.  If instead a segmented detector were used, the deviation from the rigid-intensity-shift model would be hard to gauge from the STEM images alone.  However, this loss of sensitivity to fine intensity redistribution may not necessarily be a great limitation: if the intensity redistribution is sufficiently localised within the detector segments then a rigid-intensity-shift analysis applied to segmented detector DPC-STEM may be a good approximation.  To explore this, we now compare the true phase gradient of Fig.~\ref{fig:DDDsuperr}a with that estimated by a segmented detector.

\section{Effect of intensity redistributions on segmented detector DPC-STEM accuracy}
\label{sec:segdet}

Assuming a rigid-disk-shift model, segmented detector STEM images can be used for 
quantitative phase gradient measurement via a calibration establishing the correspondence between the signal in the various detectors and the magnitude and direction of the disk deflection.  Majert and Kohl \cite{majert2015high} present  analytic expressions for deflected bright-field disk overlap with detector segments.  Alternatively, the calibration can be carried out experimentally \cite{SFSMSKOMI1,zweckdetector,schwarzhuber2017achievable,brownmeasuring}, which has the advantage of accounting for the realistic detector response and some spreading of the bright-field disk via inelastic scattering.

\begin{figure}[ht]
\centering
\includegraphics[width=0.95\columnwidth]{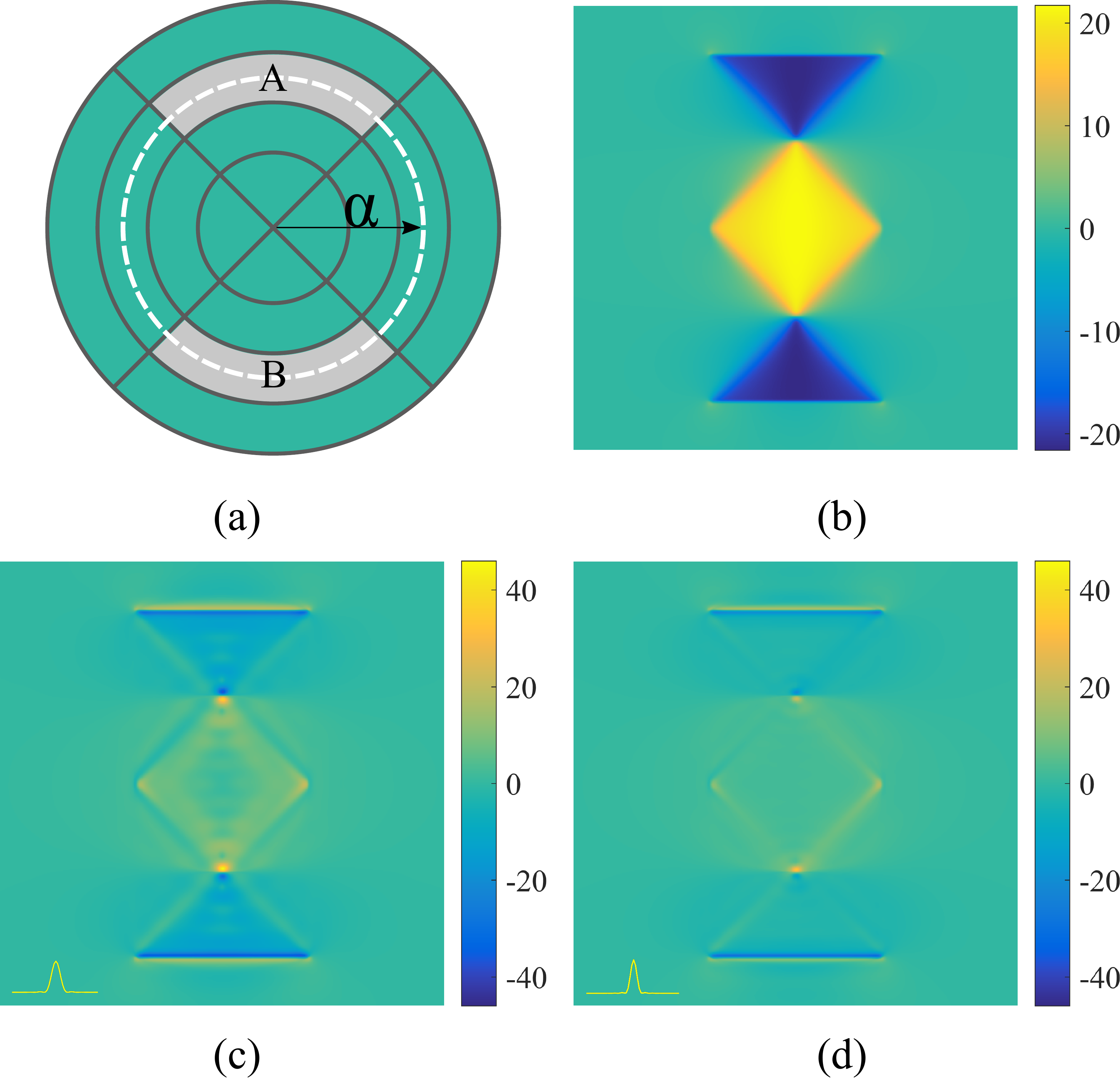}
\caption{(a) Illustration of location of the edges of the off-sample bright-field disk (white, dashed line) compared to the segmented detector. (b) $\partial \phi / \partial y$ of the phase profile in Fig.~\ref{fig:DDDsuperr}. Colour bar is in terms of probe deflection angle, $\beta$, in $\mu$rad. Difference between the calibrated segmented detector DPC-STEM estimate for $\partial \phi / \partial y$ and the true value for convergence semiangles of (c) $90\, \mu$rad and (d) $133\, \mu$rad, with the corresponding probe intensity profiles plotted to scale in the lower left corners.  Colour bars for (c) and (d) given in terms of percentage error from maximum $\beta$ in (b).}
\label{DIFFfig}
\end{figure}

Our simulations were set up as follows. To emulate the experimental set-up used to obtain Fig.~\ref{fig:DPCresolution}(d), the detector was oriented as indicated in Fig.~\ref{DIFFfig}(a), with camera length chosen such that the bright-field disk extends to midway through the third ring of detector segments.  To calibrate phase measurements of $\partial \phi / \partial y$, for each convergence semiangle considered a look-up table was generated relating the difference in intensity recorded in segment A and segment B (see Fig.~\ref{DIFFfig}(a)) to the actual bright-field disk shift.  Note that a more elaborate approach, perhaps based on approximate centre of mass, would be needed to handle deflections which shift the bright-field disk completely off either of segments A or B.  Fig.~\ref{DIFFfig}b shows $\partial \phi / \partial y$ for the magnetic domain structure of Fig.~\ref{fig:DDDsuperr}(a).  For convergence semiangles of $90\ \mu$rad and $133\ \mu$rad respectively, Figs.~\ref{DIFFfig}c and d show the difference between the calibrated segmented detector DPC-STEM estimate for $\partial \phi / \partial y$ and the true value.  

The segmented detector measurement of the phase gradient is least accurate near to regions of phase gradient change (most significantly at the top and bottom edges of the sample).  However, in Fig.~\ref{DIFFfig}c further differences are perceptible as more subtle ripples lying horizontally across the domains in the difference image. These are regions where the true phase gradient is linear, but peaks and troughs in the diffraction patterns pass on and off the detector segments, as a result of the broad probe tails. The dynamics of this rippling behaviour are shown in more detail in supplementary video A.

In Fig.~\ref{DIFFfig}d, the differences are more localised to regions of strong phase gradient change.  As previously seen in Fig.~\ref{fig:blerghhh}, changing the convergence semiangle does not necessarily alter the angular extent of the intensity redistribution at the edges of the bright-field disk.  However, because we assume camera lengths such as to maintain the same geometric overlap between the detector segments and the bright-field disk, as the convergence semiangle increases the intensity redistribution on the edges of the bright-field disk become more localised with respect to the detector segments.  This is shown in supplementary video B.

Fig.~\ref{DIFFfig}c and d show the typical error is of the order of $10\%$ of the signal.  The largest errors are strongly localised to specific features.  For most of the imaged area the errors remain small: segmented detector DPC-STEM can give good quantitative results even when, as shown in Sec.~\ref{pnsection} and  \ref{sssec:DDD}, there are significant deviations away from a rigid disk shift.

\section{Effect of probe shaping on segmented detector DPC-STEM accuracy}\label{suggestion}

The analytic modelling in Sec.~\ref{pnsection} and the ripples in the difference map in Fig.~\ref{DIFFfig}(c) suggest that much of the remaining discrepancies are  attributable to the long probe tails. It follows that if the interrogating probe can be reshaped to minimise the breadth and intensity of the probe tails then the accuracy of quantitative segmented detector DPC-STEM would improve further still.

Novel electron probe shaping has become feasible over the last few years, primarily in conjunction with studies into electron vortex beams \cite{bliokhtheory, verbeeck2010, mcmorran2011, lloydelectron, mcmorranorigins}. A number of routes to shape electron probes were developed, including manipulating optical aberrations \cite{laura, tim}, exploiting the mean inner potential of materials \cite{shiloh, beche} and using nanoscale magnetic fields \cite{bechemagnetic, blackburnvortex}.  In particular, as it is now possible to produce probes that do not have the long probe tails of the Airy-probe, we investigate the effect of reduced probe tail width on the quantitative accuracy of segmented detector DPC-STEM.

A simple probe shape with reduced tail intensity is a Gaussian probe, cf. Fig.~\ref{fig:PlotAiryGauss}. The literature gives two different routes to creating such a probe. Recent work by McMorran \emph{et al.} has used electron phase plates to form a Gaussian wavefront directly \cite{McMorranFEMMS}.  A less elegant method -- but one perhaps simpler to employ since such phase plates are not yet widely available and inserting them into electron microscopes is non-trivial -- would be to use judicious combinations of the several condenser apertures typically available in the microscope.  A first condenser aperture would create the standard Airy-disk electron probe, and a later (but still pre-sample) aperture could be used to truncate the probe at a minimum of the Airy disk. Such probe truncation (as a simple example of apodisation) is well known in astronomy and visible light optics and can produce a good approximation to a Gaussian beam \cite{bornwolf, Apodisation}.

\begin{figure}[ht]
\centering
\subfloat[][]{
\includegraphics[width=0.9\columnwidth]{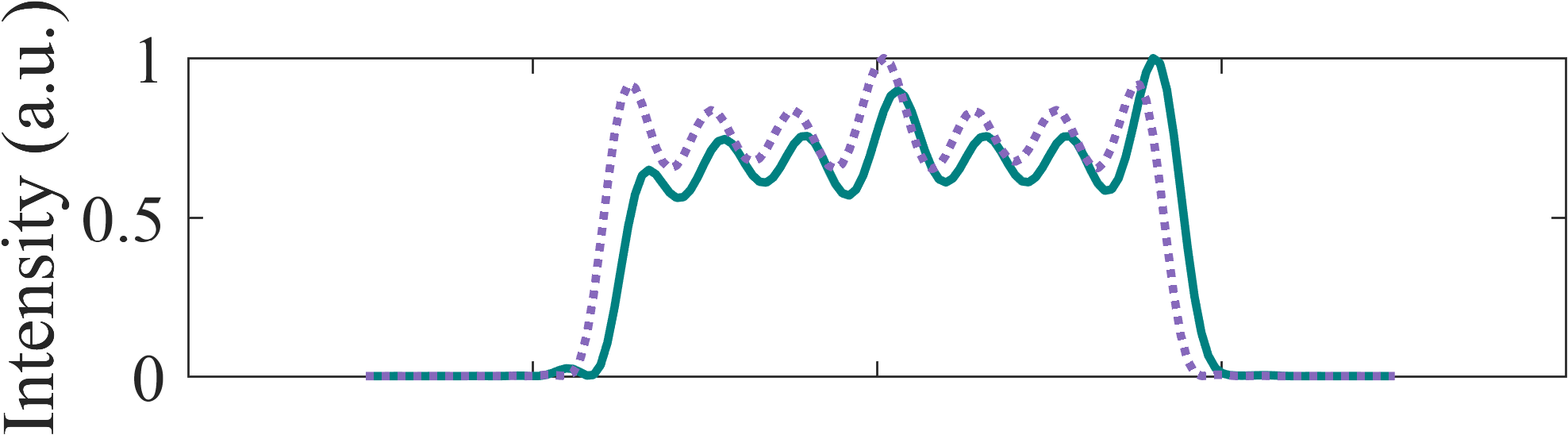}
\label{fig:subfigq1}}\\[2ex]
\subfloat[][]{
\includegraphics[width=0.9\columnwidth]{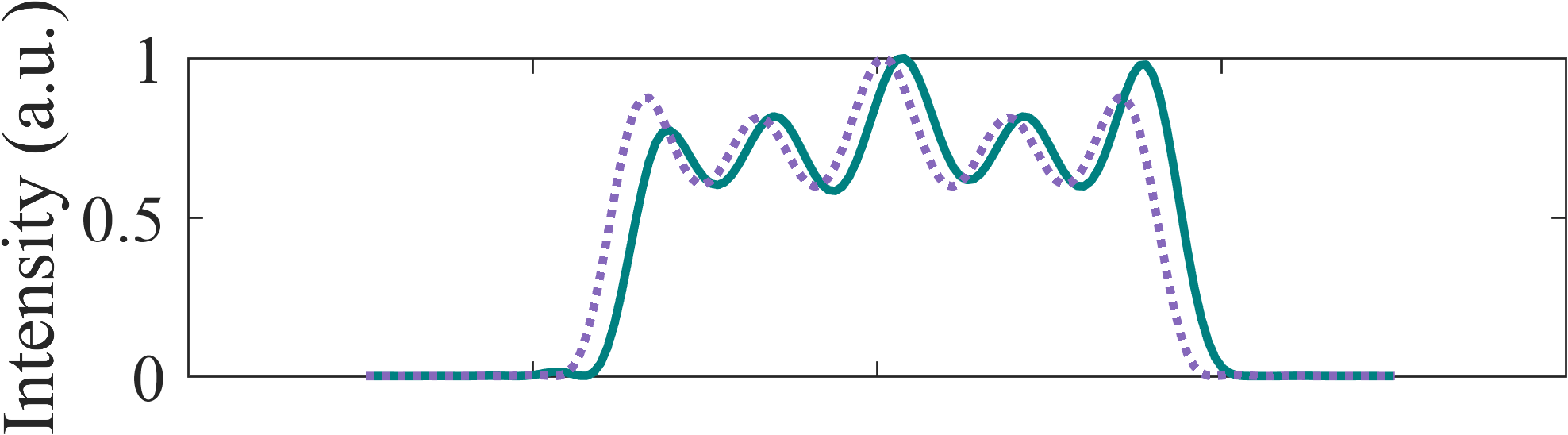}
\label{fig:subfigq2}}\\[2ex]
\subfloat[][]{
\includegraphics[width=0.9\columnwidth]{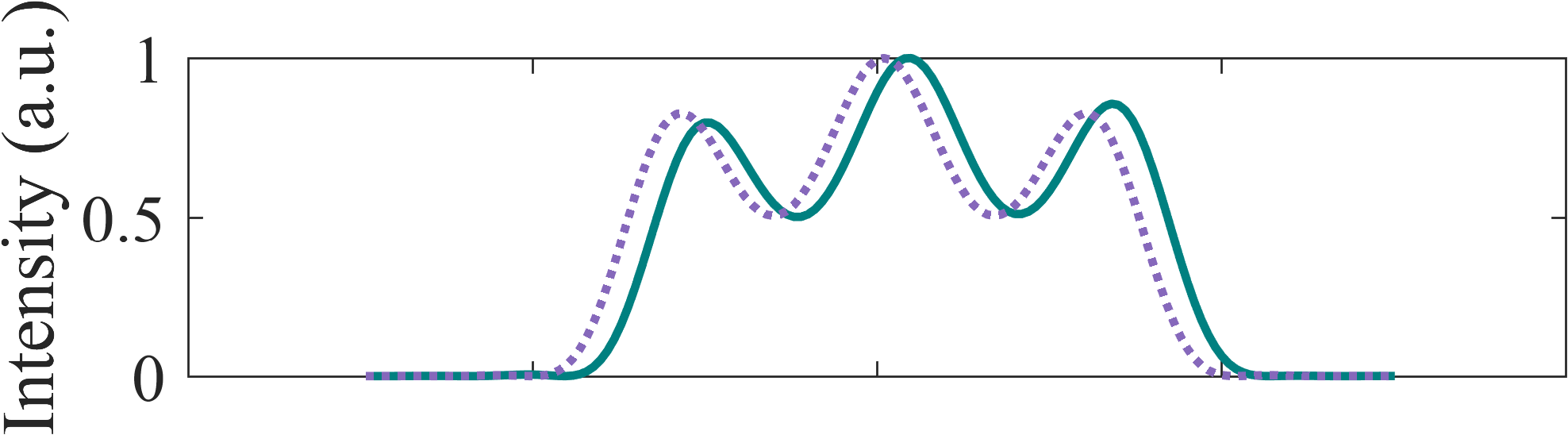}
\label{fig:subfigqq2}}\\[2ex]
\subfloat[][]{
\includegraphics[width=0.9\columnwidth]{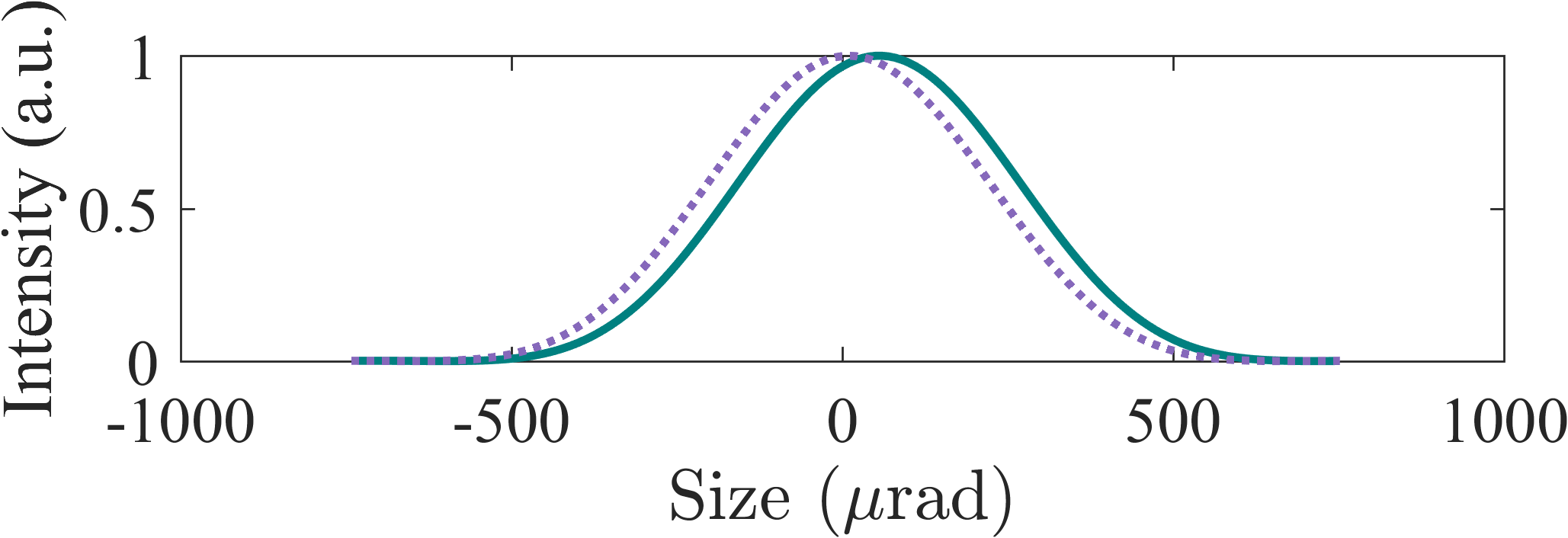}
\label{fig:subfigqq4}}\\
\caption{Line profiles of diffraction intensity from 426 $\mu$rad convergence semiangle Airy probes apodised at the (a) $7^{\rm th}$, (b) $5^{\rm th}$, (c) $3^{\rm rd}$, and (d) $1^{\rm st}$ radial minimum, for the probe off (lilac, dashed) and on (teal, solid) the p-n junction described by Eq.~(\ref{eq:pnphase}).}
\label{croppedAiry}
\end{figure}

We return to the p-n junction case, Eq.~(\ref{eq:pnphase}), to demonstrate the changes in the diffraction plane caused by successive apodisations of the Airy probe. Fig.~\ref{croppedAiry} shows simulated intensity profiles of the diffraction pattern when the probe illuminates the centre of the p-n junction and is apodised at the $7^{\rm th}$,  $5^{\rm th}$,  $3^{\rm rd}$ and  $1^{\rm st}$ Airy minima.  As compared with the top-hat diffraction intensity profile of an unapodised probe, the increasingly narrow apodisation is seen to make the profile more Gaussian (the expected form of the diffraction pattern of a Gaussian probe).  In Fig.~\ref{croppedAiry}(a), left-right asymmetry within the diffraction pattern is evident, echoing much of the behaviour of Fig.~\ref{fig:blerghhh}.  The central position of the diffraction pattern intensity may be somewhat shifted, but the intensity redistribution between the off-junction and on-junction cases is not a simple rigid shift.  However, as the apodisation radius becomes increasingly narrow, the intensity redistribution within the disc decreases in significance, and the shift of the diffraction pattern becomes clearer -- the behaviour predicted by the initial rigid-disk-shift model. In Fig.~\ref{croppedAiry}(d), the on-junction intensity is a simple shifted version of the off-junction intensity. 

\begin{figure*}[htb!]
\includegraphics[width=0.9\textwidth]{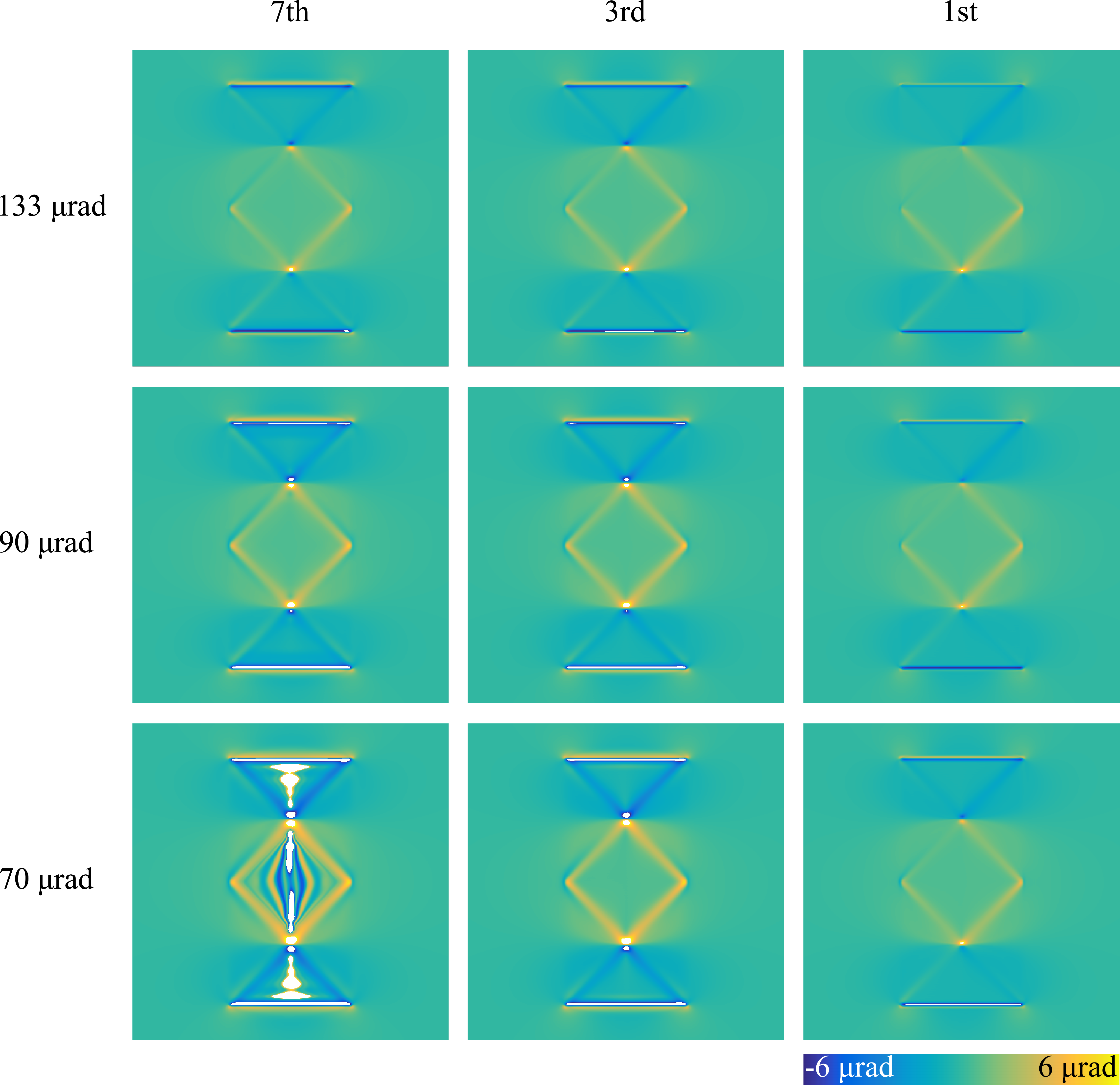}
	\caption{Difference maps in deflection angle between the true phase gradient and that calculated with the look-up table, for different $\alpha$ (rows) and apodisation (columns). The differences decrease with increasing $\alpha$, as expected, but decrease more strongly with increasingly narrow apodisation: the $1^{\rm st}$ Airy apodisation cases (right-hand column) all giving similar difference measurements. The colour maps are restricted to $\pm 6\ \mu$rad to visualise the details more clearly. Difference values outside this range have been set to white. \label{fig:ApodComp}}
\end{figure*}

To explore whether this probe shaping improves the accuracy of quantitative segmented detector DPC-STEM, we turn again to the magnetic domain case study of Sec.~\ref{sssec:DDD} and \ref{sec:segdet}.  Over a series of convergence semiangles ($70\, \mu$rad, $90\, \mu$rad and $133\, \mu$rad) and apodisation cutoffs (at the $7^{\rm th}$, $3^{\rm rd}$ and $1^{\rm st}$ Airy minima), segmented detector simulations were performed to find the phase gradients obtained.  The difference between these measured and true phase gradients, depicted in Fig. \ref{fig:ApodComp}, show increasing localisation of the sample regions that are not accurately reconstructed -- the tails have a decreasing effect as the apodisation strength is increased.

The reconstructed phases for the probes apodised at the first Airy minimum closely match the true phase gradient, aside from when the probe is within $r_{\rm probe}$ of a strong change in phase gradient.  Reducing probe tails does indeed seem to be a promising way to improve the quantitative accuracy of segmented detector DPC-STEM.

\section{Conclusion}

In this study, the break down of the rigid-intensity-shift model of DPC-STEM when the gradient of the imparted phase varies across the incident wavefunction, previously anticipated in principle and explored in the atomic resolution regime \cite{muller2014atomic,lazic2016phase,muller2017measurement,cao2017theory}, has been explored in detail for imaging long-range fields. Combining experimental, analytic modelling and simulations, we have shown that the breaking of this model is quite generic, and occurs for a range of specimen and probe parameters.  It is worth stressing that this occurs in simple phase objects; it does not require dynamical scattering.

Whether applying rigid-disk-shift interpretation is valid depends on the relationship between properties of the specimen, detector and probe.  Our conclusions can be summarised as follows:

(A) The diffraction pattern intensity redistribution will not be well described in detail by a rigid-disk-shift interpretation, irrespective of probe size, unless the product of the phase gradient and the length over which it is constant is sufficiently large (see Eq.~(\ref{eq:13})).

(B) If the convergence semiangle is sufficiently large compared to the feature size, the diffraction pattern intensity redistribution may nevertheless be confined within a detector segment, allowing an accurate, quantitative DPC-STEM reconstruction using the simple rigid-disk-shift model.

(C) As deviations from the rigid-disk-shift model are exacerbated by the broad tails of Airy probes, probe reshaping to reduce these tails can enable quantitative, accurate DPC-STEM reconstruction for a broader range of specimens.

If the diffraction patterns were recorded on a pixel detector, the break down of the rigid-intensity-shift model is not particularly problematic and may even be used to extract information about the structure.  However, that approach produces enormous data sets (on the order of $\sim 1024 \times 1024$ data points per probe position) and requires complicated analysis.  Our results show that judicious use of convergence angle and probe shaping enables quantitative, accurate phase reconstruction to be achieved using STEM images recorded on just a few detector segments.  This permits faster data collection and produces datafiles of easily manageable size, which is highly attractive for high-throughput practical applications.

\begin{acknowledgments}
This research was supported by the Australian Research Council Discovery Projects funding scheme (Project DP160102338). N.S. acknowledges support from SENTAN, JST and JSPS KAKENHI Grant numbers JP26289234 and JP17H01316.
The GaAs p-n junction samples were provided by Hirokazu Sasaki, Furukawa Electric Co., Ltd.
\end{acknowledgments}

\bibliographystyle{apsrev4-1}
\bibliography{Clark_STEMbib}
\end{document}